\documentclass[aps,prb,reprint,superscriptaddress,twocolumn]{revtex4-2}
\usepackage{amsmath}
\usepackage{graphicx}
\usepackage{dcolumn}
\usepackage{bm}
\usepackage{hyperref}
\usepackage{float}
\usepackage{soul}
\usepackage{tabularx}
\usepackage{booktabs}
\usepackage{multirow} 
\usepackage{longtable} 
\usepackage{epstopdf}
\usepackage{datetime}
\usepackage{array} 
\usepackage{makecell}
\usepackage{threeparttable} 
\usepackage{mathrsfs}
\usepackage{dsfont}
\usepackage{amssymb}
\usepackage{color}
\usepackage{ulem}

\begin{document}

\title{Thermodynamics of Classical One-dimensional Klein-Gordon Lattice Model}

\author{Hu-Wei Jia}
\affiliation{Department of Physics, Renmin University of China, Beijing 100872, P. R. China}
\affiliation{Key Laboratory of Quantum State Construction and Manipulation (Ministry of Education), Renmin University of China, Beijing 100872, China}
\author{Ning-Hua Tong}
\email{nhtong@ruc.edu.cn}
\affiliation{Department of Physics, Renmin University of China, Beijing 100872, P. R. China}
\affiliation{Key Laboratory of Quantum State Construction and Manipulation (Ministry of Education), Renmin University of China, Beijing 100872, China}
\date{\today}

\begin{abstract}
In this paper, we study the thermodynamical properties of the classical one-dimensional Klein-Gordan lattice model ($n \ge 2$) by using the cluster variation method with linear response theory. The results of this method are exact in the thermodynamical limit. We present the single site reduced density matrix $\rho^{(1)}(z)$, averages such as $\langle z^2 \rangle$, $\langle |z^n|\rangle$, and $\langle (z_1-z_2)^2\rangle$, the specific heat $C_v$, and the static correlation functions. We analyzed the scaling behavior and obtained the exact scaling powers of these quantities in the low and high temperaures. Using these results, we gauge the accuracy of the projective truncation approximation for $\phi^{4}$ lattice model.

\end{abstract}

\maketitle

\section{Introduction}

Many interesting phenomena in condensed matter physics are described by the classical many-body models. Examples include heat transport in low-dimensional system \cite{GMSch}, glass transition \cite{KNTH2018,ALBG2022,PGFH2001}, anomalous attraction \cite{Ise1} and screening \cite{Gabbie1} in Coulomb fluid, and condensation of DNA \cite{VAB1996,JUTO1995}, \textit{etc.} 

In order to study these phenomena with theoretical models, many statistical methods have been developed. On the numerical side, there are Monte Carlo (MC) \cite{NMSU1949} and molecular dynamics (MD) simulation \cite{BJTE1959,FPU1955} methods. On the analytical side, there are approaches such as variational approximations \cite{Netz1,LLW2016,LLLLW2015}, mode-coupling theory \cite{Wu1,Szamel1}, non-equilibrium Green's function \cite{Xu1}, and projective truncation approximation (PTA) for Green's functions \cite{PTAphi4}, {\it et al.}. But none of the above methods is exact.

Here, we focus on the one-dimensional classical Klein-Gordan (KG) lattice model \cite{NLBL2013}. It is an extension of the $\phi^{4}$ lattice model \cite{Parisi1} obtained from discretizing the continuous $\phi^4$ field theory \cite{Boyanovsky1}. Both models are frequently used in the study of phonon properties and thermal transport of low dimensional materials. Despite extensive numerical and analytical studies of the one-dimensional KG lattice model \cite{NLBL2013}, of which $\phi^4$ is a special case, the exact thermodynamical properties of this model have not been addressed. In this work, we will fill this gap by combining the cluster variation method (CVM) \cite{RKik1951-1} with the linear response theory (LRT) \cite{RKubo1957}. Due to the exactness of this method for one-dimensional lattice with nearest-neighbor coupling \cite{A-Peli2000,A-Peli2005,A-Peli2005-1}, we are able to obtain the exact thermodynamical properties of the one-dimensional KG lattice model in the thermodynamical limit.

We first give a brief review of CVM. CVM was first proposed by Kikuchi\cite{RKik1951-1} to describe the order-disorder transition in the lattice gas model. In the modern form of this theory, the Peierls-Fynmann-Bogoliubov variational free energy \cite{REP1938,Bogoliu1958,Fyn1972} is first formally expressed in terms of energy and entropy of the studied system. The entropy is then cumulant-expanded up to chosen clusters. Finally, the free energy is minimized with respect to the reduced density matrices of the chosen cluster \cite{TMor1957}. An \cite{GZAn1988} and Morita \textit{et al.} \cite{TMor1990} systematized the formalism and established general relations among quantities on sub-clusters and those on the largest chosen cluster, by using the concept of partially ordered set and Mobius inversion formula. 
Pelizzola \cite{A-Peli2000,A-Peli2005,A-Peli2005-1} proved that CVM is exact for Ising-like models with discrete variables on (1) one-dimensional lattice and stripe system, (2) tree-like lattice, (3) (square) cactus lattice (the interior of Husmi tree), and (4) square lattice with special coefficients. In particular, CVM based on nearest-neighbour two-site clusters is exact for models on one-dimensional chain with nearest-neighbor interaction and open boundary condition \cite{HJBr1971,JKPer1977}. The set of self-consistent CVM equations can be solved efficiently by the natural iteration method \cite{RKAB1992,RKik1998,RKik1998-2}. Unlike other cluster mean-field theories \cite{CMFT-J1J2,YD-CMFT}, an important advantage of CVM is that the spatial translation symmetry of original model is retained from the outset.  

CVM has been widely applied in the study of classical and quantum lattice models. For classical systems, such studies include the order-disorder transition in Ising-like models \cite{RKik1951-1,RKik1951-2,ENM1999,WMN1978,AIG2015}, the phase diagram of alloys \cite{RKik1974,MAFG1993} and multicomponent system \cite{SDG1984}. For quantum systems, CVM has been employed to study the liquid hellium \cite{TMor1957-2}, Heisenberg model \cite{TLFL1975,Mor-Ta1966}, s-d model \cite{Halow1968}, and the boson lattice gas \cite{TMor1994}. For reviews, see Refs.\cite{Mor-Ta1966}, \cite{Halow1968}, and \cite{THK1994}.
Although originally developed and mostly applied to Ising-like models with discrete variables, CVM can be applied to study statistical models with continuous variables \cite{RKAB1992,AF1994,TetMor2013}. In this work, we will use CVM to study the one-dimensional KG lattice model which has continuous displacements on lattice sites. 
 
In order to obtain the static spatial correlation function beyond the size of the chosen cluster (nearest-neighbor pairs in this work), we combine CVM with LRT. In CVM, LRT has been used to calculate correlation functions in Refs.\cite{TMor1990,ENM1999,WMN1978}. LRT has also been used in the closely related belief propagation to do probabilistic inference via massive probabilistic model \cite{Tanaka2004,Tanaka2005}. Note that our LRT here refers to the linear response to static external fields only.

This paper is arranged as follows. In section II, we introduce the one-dimensional KG lattice model and its scaling properties. In section III, we present the formulation of CVM and LRT. In section IV, the numerical data are shown and physical results are analysed. In section V, we compare CVM results with those from PTA \cite{PTAphi4} for the $\phi^4$ model. Section VI gives a summary and discussion.

\section{The Classical One-dimensional Klein-Gordon Lattice Model}

The Hamiltonian of the classical one-dimensional KG lattice model studied in this work reads
\begin{equation}
 H(K, \gamma) =\sum_{i=1}^L\frac{p_i^2}{2 m}+\sum_{i=1}^L \frac{K}{2}\left(x_i-x_{i+1}\right)^2+\sum_{i=1}^L \frac{\gamma}{n}\left|x_i\right|^n .    \label{ham}
\end{equation}
The dynamical variables $x_i$ and $p_i$ are the displacement from the equilibrium position and the momentum of atom $i$ ($1 \leq i \leq L$), respectively. $L$ is the total number of atoms along the chain. $n$ is the exponent of the on-site potential. $K$ and $\gamma$ are the strengths of the nearest-neighbour coupling and local potential, respectively. We exert the periodic boundary condition $x_{L+1}=x_1$ and $p_{L+1}=p_1$.  In this work we set $m=1$. 

This model Hamiltonian is obtained by discretizing the field theoretical Klein-Gordon model \cite{ADas2008,DKY2006} and generalizing the exponent of local potential from $4$ to an arbitrary exponent $n$. It is often used as a toy model for studying the effect of nonlinear interaction on lattice dynamics and the thermal transport in low-dimensional lattice system without continuous spatial translation invariance. Note that at $n=4$, KG model recovers the $\phi^4$ lattice model that is widely used in the study of anomalous heat transport in low dimensional systems \cite{Ballistic}.

Like the $\phi^4$ lattice model and FPU model \cite{FPU1955}, KG model with general parameter $n$ has the following scaling properties \cite{KADK2000,KADK2001}
\begin{equation}
  H(K, \gamma) = g \tilde{H}(1,1).
\end{equation}
Here,
\begin{eqnarray}
&& \tilde{H}(K, \gamma)  \nonumber \\
 && = \sum_{i=1}^L \frac{\tilde{p}_i^2}{2 m}+\sum_{i=1}^L \frac{K}{2}\left(\tilde{x}_i-\tilde{x}_{i+1}\right)^2+\sum_{i=1}^L \frac{\gamma}{n}\left|\tilde{x}_i \right|^n ,   \label{KGh}
\end{eqnarray}
and 
\begin{equation}
    g= \left( \frac{K^n}{\gamma^2}\right)^\frac{1}{n-2}.
\end{equation}
The rescaled variables $\tilde{x}_i$ and $\tilde{p}_i$ are related to the original ones by
\begin{eqnarray}
  &&  p_i = \sqrt{K}\left( \frac{K}{\gamma} \right)^{\frac{1}{n-2}} \tilde{p}_i,   \nonumber \\
  &&  x_i = \left(\frac{K}{\gamma} \right)^{\frac{1}{n-2}} \tilde{x}_i.   \nonumber \\
  &&
\end{eqnarray}

Note that the above transformation does not conserve the Poisson bracket. But for the statistical properties that can be calculated from the integration of variables only, the above transformation is useful in that the results for the Hamiltonian $H(K, \gamma)$ in Eq.(\ref{ham}) at arbitrary parameter $(K, \gamma)$ can be recovered from those for $H(K=1, \gamma=1)$ using the scaling relations. For examples, we have
\begin{equation}
    \langle |x|^m \rangle (K,\gamma,T)=\left(\frac{K}{\gamma}\right)^\frac{m}{n-2} \langle |x|^m \rangle \left(1,1, g^{-1}T \right),
\end{equation}
\begin{equation}
     C_v(K,\gamma,T)= g C_v  \left(1,1, g^{-1}T \right) ,
\end{equation}
and
\begin{equation}
    \xi (K,\gamma,T)=\xi \left(1,1,  g^{-1}T \right) .
\end{equation}
Therefore, in this paper, we only study the Hamiltonian $H(K=1, \gamma=1)$. For convenience, we write it as a sum of three parts,
\begin{equation}   \label{hamilton_1}
   H(K=1, \gamma=1) = \sum_{i=1}^L{\left[ E_{k}(p_i) + U(x_i) + V(x_i,x_{i+1}) \right] },
\end{equation}
with
\begin{eqnarray}          \label{hamred}
&& E_{k}(p) = \frac{p^2}{2 m},  \nonumber \\
&& U(x) = \frac{1}{n} \left| x \right|^n , \nonumber \\
&& V(x, y) = \frac{1}{2}\left(x-y \right)^2.
\end{eqnarray}
Here, $E_{k}(p)$ is the kinetic energy, $U(x)$ is the local potential, and $V(x, y)$ is the nearest-neighbor coupling. 
This form of Hamiltonian will be used in the derivation and computation below.

\section{Formula of CVM and LRT}
In this section, we present the detailed derivation of CVM and LRT formula for Hamiltonian Eq.(\ref{hamilton_1}).

\subsection{CVM formula }

For the sake of completeness, in this part we first summarize the basic formalism of CVM. Then we appliy it to the one-dimensional KG lattice model Eq.(\ref{hamilton_1}).
We start from the minimization principle of free energy $F$,
\begin{equation}   \label{free}
    F \leqslant F[\rho] \equiv \text{Tr} \left( \rho H \right) + k_B T \text{Tr} \left( \rho \ln \rho \right). 
\end{equation}
Here $F$ and $H$ are the free energy and Hamiltonian of the studied system, respectively. $\rho$ is the density operator.
Minimizing the functional $F[\rho]$ with respect to $\rho$ under the constraint $\text{Tr} \rho =1$ will give the density operator $\rho_{eq}= e^{-\beta H}/ \text{Tr} (e^{-\beta H} )$ and free energy $F$ of the equilibrium state. Here $\beta = 1/(k_B T)$, with $k_B$ being Boltzmann constant and $T$ the absolute temperature. Below, we will work under the natural unit $k_B=1$.

CVM is based on minimizing $F$ with respect to the reduced density operators of a chosen set of clusters. To do this, one first chooses a set of basic, mutually non-inclusive clusters. Then one approximately expresses the free energy functional $F[\rho]$ in terms of the reduced density operators $\rho_c$ of the basic clusters and their sub-clusters.  
For the short-range interacting system with Hamiltonian $H = \sum_c H_c$, we suppose the average energy $\langle H \rangle = \text{Tr}( \rho H) = \sum_c \text{Tr}_c( \rho_c H_c) $ can already be expressed as a functional of the cluster density operators $\rho_c$. One can express the negative entropy of a $L$-site system $-S_L = \text{Tr} \left( \rho \ln \rho \right)$ in terms of cluster quantities through the cumulant expansion,
\begin{equation}
    -S_L \equiv \Gamma_L = \sum_i^L \gamma_i^{(1)}+\sum_{i<j}^L \gamma_{i j}^{(2)}+\sum_{i<j<k}^L \gamma_{i j k}^{(3)} + ... . \label{entr}
\end{equation}
In the above equation, $\gamma_{i_1, i_2, ..., i_n}^{(n)}$ ($n=1,2,...,L$) is the cumulant entropy of the $n$-site cluster composed of sites $(i_1, i_2, ..., i_n)$. 

In CVM, the above expansion is truncated to the chosen set of basic clusters and their sub-clusters, under the assumption that the neglected terms are sufficiently small. For each cluster $c$ appearing in the truncated expansion, its entropy $S_c$ is related to the the cumulant entropies of its sub-clusters (including itself) via a similar expression,
\begin{equation}
    -S_c \equiv \Gamma_c = \sum_i^c \gamma_i^{(1)}+\sum_{i<j}^c \gamma_{i j}^{(2)}+\sum_{i<j<k}^c \gamma_{i j k}^{(3)} + ... . \label{entr_c}
\end{equation}
Here the expansion runs over all sub-clusters of $c$, including the cluster $c$ itself. This equation can be written down for every sub-cluster of $c$, generating a closed set of linear equations for all the involved cumulant entropies $\gamma_{i_1, i_2, ..., i_n}^{(n)}$ ($n=1,2,..., |c|$).  
By M\"{o}bius inversion, the linear equations can be solved to express the cumulant entropies $\gamma$'s in terms of $\Gamma_{c^{\prime}}$'s of clusters $c^{\prime} \subseteq c$ which can be explicitly written as functionals of the cluster density operators $\rho_{c^{\prime}}$,
\begin{equation}
   \Gamma_{c^{\prime}} = \text{Tr}_{c^{\prime}} \left( \rho_{c^{\prime}} \ln{\rho_{c^{\prime}} } \right).
\end{equation}
Putting $\gamma$'s back into Eq.(\ref{entr}), one finally expresses $-S_L$ as functional of cluster density operators. The variation of free energy $F[\rho]$ in Eq.(\ref{free}) is then carried out with respect to all the involved $\rho_c$, under the constraints 
\begin{eqnarray}    \label{eq15}
&&  \text{Tr}_c \rho_c =1;  \nonumber \\
&&  \text{Tr}_{c-c^{\prime}} \rho_c = \rho_{c^{\prime}},   \,\,\,\,\,\, (\forall c^{\prime} \subset c).
\end{eqnarray}
In the second line of Eq.(\ref{eq15}), $\text{Tr}_{c-c^{\prime}}$ is the partial trace over degrees of freedom in cluster $c$ but outside cluster $c^{\prime}$.

To apply CVM to Hamiltonian Eq.(\ref{hamilton_1}), we first separate the total free energy $F$ into the kinetic part $F_{kin}$ and the potential energy part $F_{pot}$,
\begin{eqnarray}
   && F = F_{kin} + F_{pot},   \nonumber \\
   && F_{kin} = -\frac{1}{\beta} \ln \left[ \displaystyle\prod_{i=1}^{L} \int\limits_{-\infty}^{+\infty} dp_i \, e^{-\beta E_{k}(p_i)} \right],  \nonumber\\
   && F_{pot} = -\frac{1}{\beta} \ln \left[\int ... \int\limits_{-\infty}^{+\infty}  \displaystyle\prod_{i=1}^{L} dx_i \, e^{-\beta \sum\limits_{i=1}^{L} \left[ U(x_i) + V(x_i,x_{i+1}) \right] } \right].  \nonumber \\
   && 
\end{eqnarray}
Only the potential energy part $F_{pot}$ needs to be treated by CVM. Therefore, in the rest part of the paper, we will use $H$, $F$, and $S_L$ to refer to the potential energy part of the corresponding quantities. The kinetic contribution to the total free energy will be added whenever necessary.

For the one-dimensional model with nearest-neighbor coupling, our basic cluster set is composed of all the nearest-neighbor pairs. It not only covers all the interaction terms in the Hamiltonian, but also is the minimum set that supports the exact cumulant expansion of $S_L$ for the one-dimensional chain with an open boundary condition. This is because the joint density distribution $\rho^{(L)}(x_1,x_2,...,x_L) $ for our $L$-site KG model with open boundary condition can be exactly expressed as \cite{A-Peli2005-1}
\begin{eqnarray}
&& \rho^{(L)}(x_1,x_2,...,x_L) \nonumber \\
 &&= \frac{\rho_{12}^{(2)}(x_1,x_2)\rho_{23}^{(2)}(x_2,x_3)...\rho_{L-1,L}^{(2)}(x_{L-1},x_L)}{\rho_{2}^{(1)}(x_2)\rho_{3}^{(1)}(x_3)...\rho_{L-1}^{(1)}(x_{L-1})}.    \label{fac}
\end{eqnarray}  
Here, $\rho_{i}^{(1)}(x_i)$ and  $\rho_{i,i+1}^{(2)}(x_i,x_{i+1})$ are the reduced density matrices of site $i$ and pair $(i, i+1)$, respectively. As a consequence, the cumulant expansion of entropy in Eq.(\ref{entr}) terminates rigorously at second order. 

Since our Hamiltonian Eq.(\ref{ham}) is defined with a periodic boundary condition, CVM based on neighbor-neighbor pairs is not exact for this Hamiltonian with finite chain length $L$. In the limit $L=\infty$, however, the open and periodic boundary conditions are equivalent for this system. As we will see below (Eq.(\ref{cvm_eq})), under the lattice translation invariance of Eq.(\ref{ham}), our CVM equations do not contain $L$. They can therefore be applied directly to $L=\infty$ to produce exact results for the thermodynamical limit.

With this choice of basic clusters, the average energy (interaction part only) reads
\begin{equation}    \label{trrhoH}
   \text{Tr} (\rho H) = \sum_{i=1}^{L} \left[ \langle U(x_i) \rangle + \langle V(x_i, x_{i+1}) \rangle \right],
\end{equation}
with 
\begin{eqnarray}   \label{aveUV}
  && \langle U(x_i) \rangle = \int_{-\infty}^{+\infty} dx \rho_i^{(1)}(x) U(x),   \nonumber \\
  && \langle V(x_i, x_{i+1}) \rangle = \int \int_{-\infty}^{+\infty} dx dy \, \rho_{i,i+1}^{(2)}(x,y ) V(x, y).  \nonumber \\
  &&
\end{eqnarray}
Functions $U(x)$ and $V(x, y)$ are given in Eq.(\ref{hamred}).
The entropy term is obtained as
\begin{eqnarray}   \label{entr_pair}
    -S_L &\equiv & \Gamma_L = \sum_{i=1}^L \gamma_i^{(1)}+\sum_{i=1}^L \gamma_{i, i+1}^{(2)}  \nonumber \\
         &=& \sum_{i=1}^L \Gamma_i^{(1)}+\sum_{i=1}^L \left[\Gamma_{i,i+1}^{(2)}-\Gamma_i^{(1)}-\Gamma_{i+1}^{(1)} \right].
\end{eqnarray}
Here, $\Gamma_i^{(1)}$ and $\Gamma_{i,i+1}^{(2)}$ are the negative entropies of single-site and n.n. pair, respectively,
\begin{eqnarray}    \label{gamma}
\Gamma_i^{(1)} &=& \int_{-\infty}^{+\infty}  d x \rho_{i}^{(1)}(x) \ln \rho_{i}^{(1)}(x),  \nonumber \\ 
\Gamma_{i,i+1}^{(2)} & =& \int \int_{-\infty}^{+\infty}  dx dy \, \rho_{i,i+1}^{(2)} (x, y) \ln \rho_{i,i+1}^{(2)} (x, y). \nonumber \\
&&                         
\end{eqnarray}

Considering the spatial translation invariance, we obtain the final expression for $F/L$ as
\begin{equation}
    F/L = \langle U(x)\rangle + \langle V(x,y) \rangle + T \left[ \Gamma^{(2)}-\Gamma^{(1)} \right].  \label{ave_form}
\end{equation}
The averages are readily expressed in terms of $\rho^{(1)}(x)$ and $\rho^{(2)}(x, y)$ as
\begin{eqnarray}      \label{ave_UV}
&& \langle U(x)\rangle = \frac{1}{n} \int_{-\infty}^{+\infty} d x \rho^{(1)}(x) \left| x \right|^n ,  \nonumber \\
&& \langle V(x,y)\rangle = \frac{1}{2} \int \int_{-\infty}^{+\infty} d x d y \, \rho^{(2)}(x, y) (x-y)^2. \nonumber \\
&&    
\end{eqnarray}
The last two terms in Eq.(\ref{ave_form}) read
\begin{eqnarray}     \label{gamma_NN}
\Gamma^{(1)} &=& \int_{-\infty}^{+\infty}  d x \rho^{(1)}(x) \ln \rho^{(1)}(x),  \nonumber  \\ 
\Gamma^{(2)} & =& \int \int_{-\infty}^{+\infty}  d x d y \, \rho^{(2)}(x, y) \ln \rho^{(2)}(x, y).  
\end{eqnarray}
Eqs.(\ref{ave_form}-\ref{gamma_NN}) give the explicit expression for the functional $F[\rho^{(1)}, \rho^{(2)}]$.

The variation of $F[\rho^{(1)}, \rho^{(2)}]$ must be carried out under the following normalization and reducibility constraints,
\begin{equation}\begin{aligned}   \label{constraint}
  &\int_{-\infty}^{+\infty}  dx \rho^{(1)}(x) = 1,    \\
  &\int_{-\infty}^{+\infty}  dy\rho^{(2)}\left(x,y\right)  = \rho^{(1)}\left(x\right),  \\
  &\int_{-\infty}^{+\infty}  dx \rho^{(2)}\left(x, y\right) = \rho^{(1)}\left(y\right),   \\
&\int \int_{-\infty}^{+\infty}  d x d y \, \rho^{(2)} \left(x, y\right) = 1.
\end{aligned}\end{equation}
Enforcing the four constraints by Lagrangian multipliers $\lambda_1$, $\lambda_2(x$), $\lambda_3(x)$, and $\lambda_4$, respectively, we obtain the functional to be varied as
 \begin{widetext}
 \begin{eqnarray}    \label{free_v}
\tilde{F}/L &=& F/L  - \lambda_{1}\left[\int_{-\infty}^{+\infty}  d x \rho^{(1)}(x)-1\right]
- \int_{-\infty}^{+\infty}  dx \lambda_{2} (x) \left[\int_{-\infty}^{+\infty}  d y \rho^{(2)}(x, y)-\rho^{(1)}(x)\right]    \nonumber \\
&& - \int_{-\infty}^{+\infty}  dy \lambda_{3}(y) \left[\int_{-\infty}^{+\infty}  d x \rho^{(2)}(x, y)-\rho^{(1)}(y)\right] 
- \lambda_{4}\left[\int \int_{-\infty}^{+\infty} d x d y \, \rho^{(2)}(x, y)-1 \right].
\end{eqnarray}
\end{widetext}

Carrying out the functional derivative of $\tilde{F}/L$ with respect to $\lambda_1$, $\lambda_2(z)$, $\lambda_3(z)$, and $\lambda_4$ gives the four constraints Eq.(\ref{constraint}), while with respect to $\rho^{(1)}(z)$ and $\rho^{(2)}(z,w)$ gives
\begin{equation}
    \rho^{(1)}(z)=e^{\beta \left[U(z) + \lambda_{2}(z) + \lambda_{3}(z)- \lambda_{1} \right]-1},   \label{rho1}
\end{equation}
and
\begin{equation}
 \rho^{(2)}(z, w)=e^{\beta \left[-V(z,w) +\lambda_{2}(z)+\lambda_{3}(w) + \lambda_{4} \right] -1}.    \label{rho2}
\end{equation}
For the case of symmetric coupling $V(x-y) = V(y-x)$ studied in this paper (See Eq.(\ref{hamred})), we can prove that $\lambda_{2}(z) = \lambda_3(z)$, which can be determined only up to an additive constant. Therefore, the one-site reduced density matrix $\rho^{(1)}(z)$ can be rewritten as
\begin{equation}
	\rho^{(1)}(z) = e^{\beta \left[ U(z) + 2 \lambda_2(z) - \lambda_1 \right] -1}.   \label{rho1_a}
\end{equation}
Similarly, $\rho^{(2)}(z,w)$ is given by
\begin{equation}
	\rho^{(2)}(z, w)=e^{\beta \left[-V(z,w) +\lambda_{2}(z)+\lambda_{2}(w) + \lambda_{4} \right] -1}.    \label{rho1_b}
\end{equation}

Substituting Eq.(\ref{rho1_a}) and Eq.(\ref{rho1_b}) into Eqs.(\ref{constraint}), the self-consistent equations about $\lambda_1 + \lambda_4$ and $\lambda_2(z)$ is finally reduced to the following form,
\begin{eqnarray}    \label{cvm_eq}
&&   e^{\beta (\lambda_1 + \lambda_4)} = \frac{\int_{-\infty}^{+\infty} dz  \, e^{\beta \left[U(z) + 2 \lambda_2(z) \right] }}{\int_{-\infty}^{+\infty} dz \, e^{\beta \left[Q(z) + \lambda_2(z) \right] }},   \nonumber \\
&& \lambda_2(z) = Q(z) - U(z) +    \lambda_1 + \lambda_4,  \nonumber \\
&& e^{\beta Q(z) } = \int_{-\infty}^{+\infty} dw \,  e^{\beta \left[ \lambda_2(w) - V(z,w) \right] }.
\end{eqnarray}
The above formula of CVM for one-dimensional KG lattice model does not contain the system size $L$ and hence applies to arbitrary size, in particular, $L=\infty$. In this limit, the periodic boundary condition used in our Hamiltonian is equivalent to the open boundary condition for which the two-site CVM is exact. The numerical confirmation of this issue is presented in Fig.\ref{Fig7}.

\subsection{Linear Response Theory}

The above CVM formalism only produces $\rho^{(1)}(z)$ and $\rho^{(2)}(z,w)$, which contains local and nearest-neighbor correlations. In order to obtain the long-range correlation functions such as $\langle x_i x_j \rangle$ for arbitrary $i-j$, we resort to the linear response theory.

We add the following local external field terms to the original Hamiltonian Eq.(\ref{hamred})
\begin{equation}
    H^{\prime}= \sum_{i=1}^{L} h_i B(x_i).
\end{equation}
Here $h_i$ is the local field on site $i$ and $B(x_i)$ is a function of $x_i$, to be specified below. The total Hamiltonian at $K=1$ and $\gamma=1$ becomes
\begin{equation}
H = \sum_{i=1}^L \frac{p_i^2}{2 m}+\sum_{i=1}^L \frac{1}{n}\left|x_i\right|^n + \sum_{i=1}^L \frac{1}{2}\left(x_i-x_{i+1}\right)^2 + H^{\prime}.    \label{KGex}
\end{equation}
Applying the neighbor-neighbor pair CVM to the above $H$, we obtain the variational free energy (interaction part)
\begin{eqnarray}
    F &=&  \sum_{i=1}^{L}\langle U(x_i)\rangle + \sum_{i=1}^{L} \langle V(x_i,x_{i+1}) \rangle
          + \sum_{i=1}^{L} h_i\langle B(x_i) \rangle   \nonumber \\
          && + T  \sum_{i}^L \left[\Gamma_{i,i+1}^{(2)}- \Gamma_i^{(1)}\right].   \label{fr_en}
\end{eqnarray}
Here, $\langle U(x_i) \rangle$, $\langle V(x_i,x_{i+1})\rangle $ are given by Eq.(\ref{aveUV}) and $\Gamma_i^{(1)}$, $\Gamma_{i,i+1}^{(2)}$ by Eq.(\ref{gamma}).

In the above equation, $\rho^{(1)}_i(x)$ and $\rho^{(2)}_{i,i+1}(x,y)$ obey the following constraints, 
\begin{eqnarray}    \label{constraint_loc}
  && \int_{-\infty}^{+\infty}  dx \, \rho_{i}^{(1)}(x) = 1,    \nonumber \\
  && \int_{-\infty}^{+\infty}  dy \, \rho_{i,i+1}^{(2)}\left(x,y\right)  = \rho_{i}^{(1)}\left(x\right),  \nonumber \\
  && \int_{-\infty}^{+\infty}  d x \, \rho_{i,i+1}^{(2)}\left(x, y\right) = \rho_{i+1}^{(1)}\left(y\right),   \nonumber \\
&& \int \int_{-\infty}^{+\infty}  d x d y  \, \rho_{i,i+1}^{(2)} \left(x, y\right) = 1.
\end{eqnarray}
They are enforced by Lagrangian multipliers $\lambda_{1i}$, $\lambda_{2i}(z)$, $\lambda_{3i}(z)$, and $\lambda_{4i}$, respectively. We get
\begin{eqnarray}   \label{v_fr_en} 
\tilde{F} &=& F - \sum_{i=1}^L \lambda_{1i} \left[\int_{-\infty}^{+\infty} d x \rho_{i}^{(1)}(x)-1 \right] \nonumber \\
&& - \sum_{i=1}^L \int_{-\infty}^{+\infty} dx \lambda_{2i}(x) \left[\int_{-\infty}^{+\infty} d y \rho_{i,i+1}^{(2)}(x,y) 
-\rho_{i}^{(1)}(x)\right] \nonumber \\
&&  -\sum_{i=1}^{L} \int_{-\infty}^{+\infty} dy \lambda_{3 i}(y) \left[\int_{-\infty}^{+\infty} d x \rho_{i,i+1}^{(2)} (x, y)-\rho_{i+1}^{(1)}(y) \right]   \nonumber \\
&& -\sum_{i=1} \lambda_{4i} \left[\int \int_{-\infty}^{+\infty} d x dy \, \rho_{i,i+1}^{(2)}(x,y)-1 \right].           
\end{eqnarray}

Carrying out the variation of $\tilde{F}$ with respect to $\lambda_{1i}$, $\lambda_{2i}(z)$, $\lambda_{3i}(z)$, and $\lambda_{4i}$ will produce the constraints equations Eqs.(\ref{constraint_loc}). The variation with respect to $\rho^{(1)}_{i}(z)$ and $\rho^{(2)}_{i,i+1}(z,w)$ then gives
\begin{equation} \label{rho1_loc}
    \rho_{i}^{(1)}(z)= e^{\beta \left[ U(z)- h_i B(z) + \lambda_{2i}(z) + \lambda_{3i}(z) - \lambda_{1i} \right] -1},
\end{equation}
and
\begin{equation} \label{rho2_loc}
  \rho_{i,i+1}^{(2)}(z, w) = e^{\beta \left[ -V(z,w) + \lambda_{2i}(z)+ \lambda_{3 i+1}(w) + \lambda_{4i} \right]-1}.
\end{equation}
In the limit $\{ h_i \} =0$, the above equations reduce to the uniform CVM equations Eqs.(\ref{rho1}) and (\ref{rho2}).

To obtain the correlation function, we calculate the differentiation of the above CVM equations with respect to field $h_1$, and then take the limit $h_i =0$ for all $i$. We define the response of the Lagrangian multipliers $\lambda_{1i} \sim \lambda_{4i}$ to $h_1$ as
\begin{eqnarray}
&&   \chi_{1i} \equiv \frac{\partial \lambda_{1i}}{\partial h_1} \Big|_{ \{h_i\} =0},  \nonumber\\
&&   \chi_{2i}(z) \equiv \frac{\partial \lambda_{2i}(z)}{\partial h_1} \Big|_{ \{h_i\} =0},  \nonumber\\
&&   \chi_{3i}(z) \equiv \frac{\partial \lambda_{3i}(z)}{\partial h_1} \Big|_{ \{h_i\} =0},  \nonumber\\
&&   \chi_{4i} \equiv \frac{\partial \lambda_{4i}}{\partial h_1} \Big|_{ \{h_i\} =0}.   
\end{eqnarray}
Substituting Eqs.(\ref{rho1_loc}) and (\ref{rho2_loc}) into (\ref{constraint_loc}), doing the derivative of $h_1$, and taking the limit $h_i=0$ for all $i$, we obtain a set of self-consistent equations about $\chi_{1i}$, $\chi_{2i}(z)$, $\chi_{3i}(z)$, and $\chi_{4i}$, where the CVM uniform solutions for $\lambda_1$, $\lambda_2(z)$, $\lambda_4$, and $\rho^{(1)}(z)$ are used as input. The equations can be simplified in the momentum space by introducing the Fourier transform of $\chi_{1i}$, $\chi_{2i}(z)$, $\chi_{3i}(z)$, and $\chi_{4i}$,
\begin{equation}
   \chi_{\alpha k} \equiv \sum_{j=1}^{L} \chi_{\alpha j} e^{-i k(j-1)},   \,\,\,\,\,(\alpha=1,2,3,4).
\end{equation}
Here, $k = 2\pi m /L$ ($m=0,1,..., L-1$).
The self-consistent equations for $\chi_{\alpha k}$ ($\alpha=1 \sim 4$) is finally obtained as
\begin{eqnarray}      \label{lrt_eq}
   \chi_{1k} &=& \int_{-\infty}^{\infty} dz \, \rho^{(1)}(z) \left[\chi_{2k}(z) + \chi_{3k}(z) - B(z) \right], \nonumber \\
  \chi_{4k} &=& - \int\int_{-\infty}^{+\infty} dz dw \, \rho^{(2)}(z,w) \left[\chi_{2k}(z) + e^{ik} \chi_{3k}(w) \right], \nonumber \\
  \chi_{2k}(z) &=&  B(z) + \chi_{1k}  +  e^{-ik}\chi_{4k}\nonumber \\
     &&  + \, e^{-ik} \left[ \int_{-\infty}^{+\infty} dw \, F(w,z)\chi_{2k}(w)\right], \nonumber \\
 \chi_{3k}(z) &=& B(z) + \chi_{1k} +\chi_{4k} \nonumber \\
    &&  + \, e^{ik} \left[ \int_{-\infty}^{+\infty} dw \, F(w,z)\chi_{3k}(w)\right].  
\end{eqnarray}
In the above equations, the function $F(w,z)$ is given by
\begin{equation}     \label{fwz}
  F(w,z) = e^{\beta \left[-V(w,z) + \lambda_2(w) - Q(z) \right]}.
\end{equation}
Here, $\lambda_2(w)$ and $Q(z)$ are solved from Eq.(\ref{cvm_eq}).

After Eq.(\ref{lrt_eq}) is solved to produce $\chi_{1k} \sim \chi_{4k}$, the correlation function $C_{1i}\equiv  \langle B(x_1) B(x_i) \rangle -\langle B(x_1) \rangle \langle  B(x_i) \rangle$ can be calculated as
\begin{eqnarray}    \label{c1i_def}
  && C_{1i} \nonumber \\
  &&  =  \frac{1}{\beta} \frac{\partial \langle B(x_i) \rangle}{ \partial f_1}\Big|_{\{f_i =0 \}}  \nonumber \\
         && = \int_{-\infty}^{+\infty} dz \, \rho^{(1)}(z) B(z) \times   \nonumber \\ 
 && \,\,\,\,\,\,\,\,\,\,\,\, \,\,\,\,\,\,\,\,\,\,\,\,\, \,\,\,\,\, \left[ -\delta_{1,i}B(z) - \chi_{1i} + \chi_{2i}(z) + \chi_{3i}(z) \right]. \nonumber \\
&&         
\end{eqnarray} 
The corresponding formula in momentum space reads
\begin{eqnarray}    \label{ck}
   && C(k)    \nonumber\\
   &&=\int_{-\infty}^{\infty} dz \, \rho^{(1)}(z) B(z) \left[- B(z) -\chi_{1k} + \chi_{2k}(z) + \chi_{3k}(z)\right], \nonumber\\
&&   
\end{eqnarray}
which can give out $C_{1j}$ through
\begin{eqnarray}    \label{c1i_cal}
    C_{1j} = \frac{1}{L} \sum_{k} C(k) e^{ik(j-1)}.
\end{eqnarray}
In Eq.(\ref{lrt_eq}), we can assign $\chi_{3k}(z) = \chi_{2k}^{\ast}(z)$ to obtain a real $C(k)$.

Note that Eqs.(\ref{lrt_eq}) and (\ref{ck}) depend on the system size $L$ through the $k$ discretization. They are exact for the thermodynamical limit $L=\infty$ but not for a chain with finite length $L$. The same is true for Eq.(\ref{c1i_cal}).

\section{results}

In this section, we present results obtained by numerically solving the CVM and LRT equations. We also give the asymptotic expressions valid in the low and high temperature limits. The results are for parameters $K=1$ and $\gamma=1$. For the numerical integration in Eqs.(\ref{cvm_eq}) and (\ref{lrt_eq}), we use the Gauss-Hermite Quadrature based on 800 point of weights and meshes generated by Matlab software. We investigate the following quantities: (A) single-site reduced density matrix $\rho^{(1)}(z)$, (B) $\langle |z|^n \rangle$, $\langle z^2\rangle$, $\langle (z_1-z_2)^2\rangle$, (C) specific heat, and (D) static correlation function. Only (D) involves LRT calculation. 

\subsection{Single-Site Reduced Density Matrix $\rho^{(1)}(z)$}

\begin{figure}       
 \vspace*{-5.0cm}
\begin{center}
  \includegraphics[width=550pt, height=400pt, angle=0]{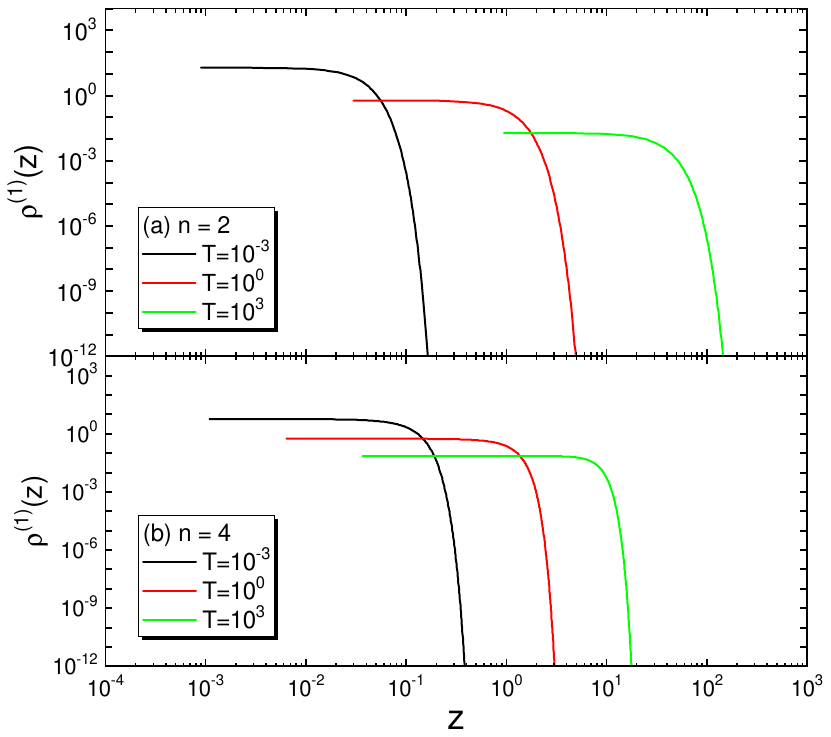}
  \vspace*{-1.0cm}
\end{center} 
\caption{(color online)
Single-site reduced density matrix $ \rho^{(1)}(z)$ as a function of $z$ for different temperatures. (a) $n=2$, and (b) $n=4$. \label{Fig1} 
}
\end{figure}
In Fig.\ref{Fig1}, we show the single-site reduced density matrix $\rho^{(1)}(z)$ in $z>0$ for two representative $n$ values, $n=2$ (harmonic oscillator) and $n=4$ ($\phi^4$ model). Due to $\rho^{(1)}(-z) = \rho^{(1)}(z)$, the $z<0$ regime is not shown. Other $n$ values give similar bell-like one-peak curves with maximum at $z=0$. With increasing temperature, the peak of $\rho^{(1)}(z)$ becomes broader and lower. Below, we will denote $\rho^{(1)}(z)$ at temperature $T$ as $\rho^{(1)}(z, T)$. For a given $n$, the height of the peak, $H(n,T) \equiv \rho^{(1)}(z=0, T)$, is a decreasing function of $T$. The shape of the curve $\rho^{(1)}(z)$ reflects the competition between the thermal movement of particles and the shape of confining potential $|z|^{n}$. For a fixed $T$, as $n$ increases, the local potential becomes flatter at bottom in $-1 \leqslant z \leqslant 1$ and increases more sharply at $z = \pm 1$. As a result, $\rho^{(1)}(z)$ has a flatter distribution at $|z| \leqslant 1$ and decreases sharply at $|z| \geqslant 1$. In the limit of $n =\infty$, the local potential becomes an infinitely deep potential well in $|z| \leqslant 1$ with a flat bottom. $\rho^{(1)}(z)$ will then have a rectangular-shaped distribution in $|z| \leq 1$.

%
\begin{figure}
 \vspace*{-0.5cm}
\begin{center}
  \includegraphics[width=300pt, height=220pt, angle=0]{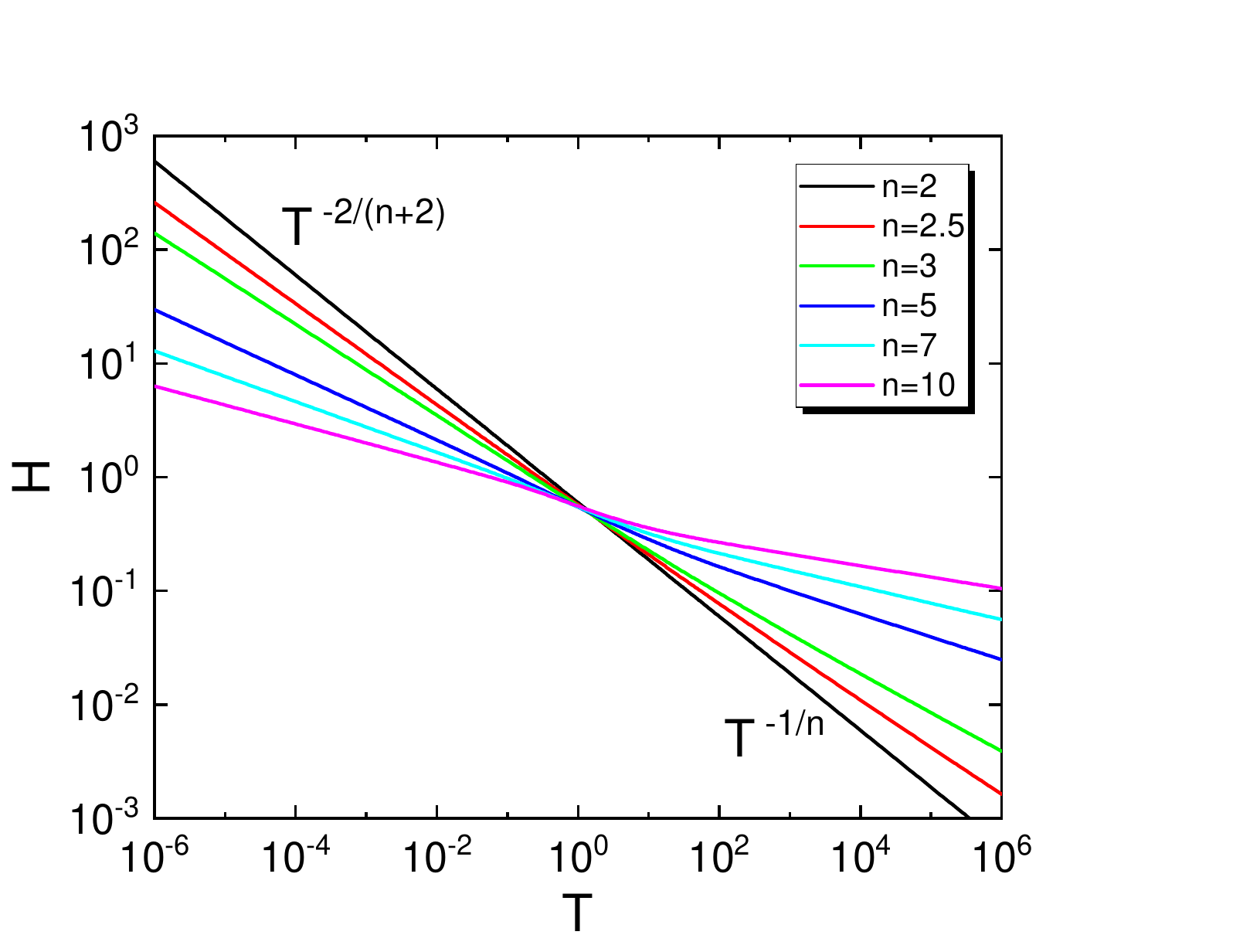}
  \vspace*{-1.0cm}
\end{center} 
\caption{(color online) The hight $H$ of $\rho^{(1)}(z)$ curve as functions of $T$ for various $n$ values. The low temperature asymptotic behavior $H \sim T^{-\frac{2}{n+2}}$ and the high temperature $H \sim T^{-\frac{1}{n}}$ are marked in the figure.
} \label{Fig2}
\end{figure} 
\begin{figure}
 \vspace*{-0.5cm}
\begin{center}
  \includegraphics[width=300pt, height=220pt, angle=0]{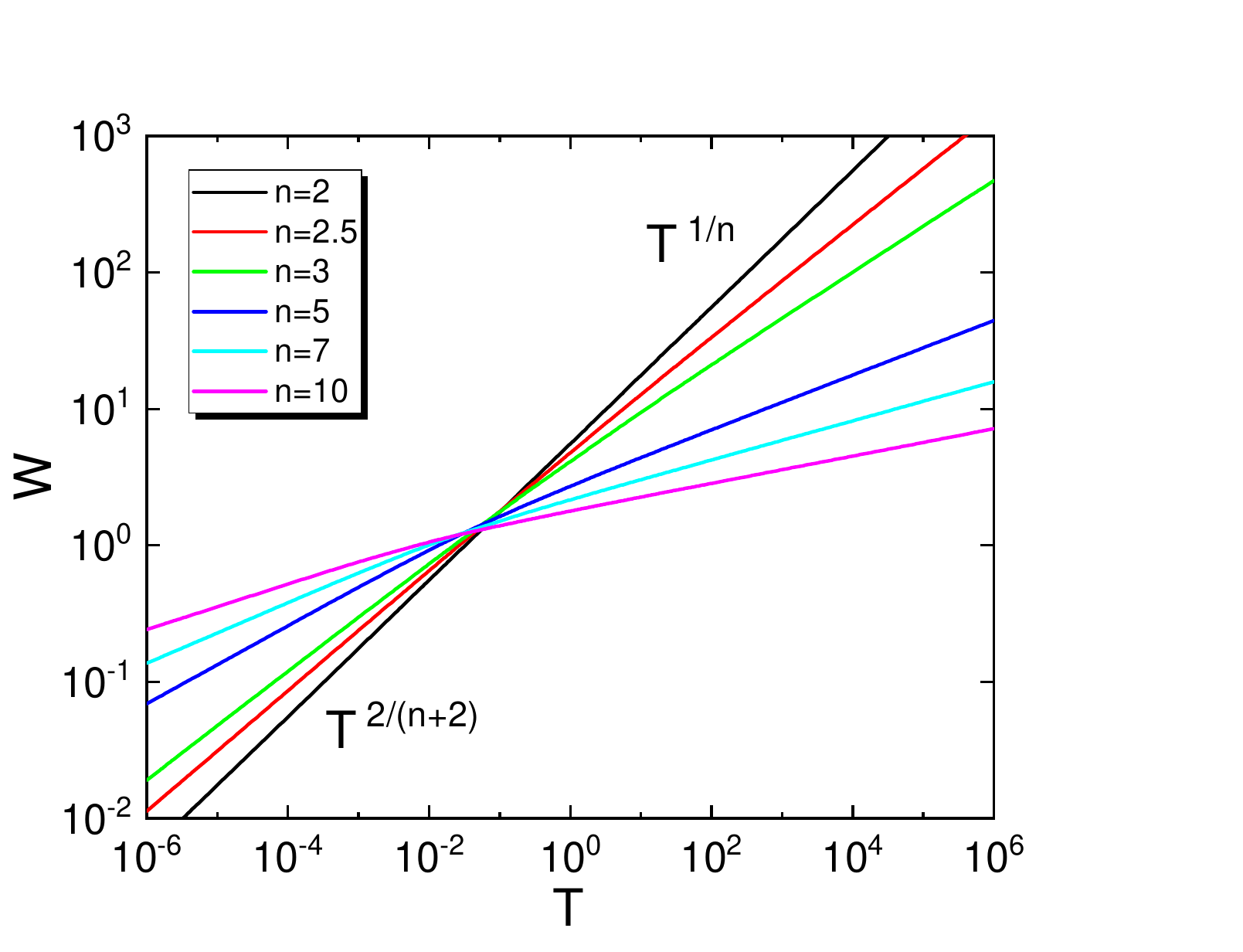}
  \vspace*{-1.0cm}
\end{center} 
\caption{(color online) Width of $\rho^{(1)}(z)$ curve as functions of $T$ for various $n$ values. The low temperature asymptotic behavior $W \sim T^{\frac{2}{n+2}}$ and the high temperature $\rho^{(1)}(z) \sim T^{\frac{1}{n}}$ are marked in the figure.
} \label{Fig3}
\end{figure}

In Fig.\ref{Fig2}, we plot the peak height $H(n,T)$ as functions of $T$ for various $n$ values. It has different low and high temperature asymptotic power laws. Fitting the numerical data gives the following form,
\begin{equation}
    H(n,T) \sim 
    \left\{\begin{array}{lll}
        T^{-\frac{2}{n+2}}, & \,\,\,\,\,\,\,\, (T\ll T_{cr}); \\ 
        &&    \\
        T^{-\frac{1}{n}}, &  \,\,\,\,\,\,\,\,  (T \gg  T_{cr}). 
     \end{array}\right.       \label{pow1}
\end{equation}
For $ 2.0 \leqslant n \leqslant 10.0$ shown in Fig.2, the crossover temperature $T_{cr} \sim 1$, without significant $n$ dependence. For $n=2$, the low and high temperature exponents coincide. We have $H(n=2, T) \sim T^{-1/2}$ in the full temperature regime.

In Fig.\ref{Fig3}, we show the peak width $W(n,T)$ of the curve $\rho^{(1)}(z)$ as functions of $T$ for different $n$ values. $W(n,T)$ is defined as twice the positive solution $z$ of the equation $\rho^{(1)}(z,T)/\rho^{(1)}(z=0,T) = c$ for $c=10^{-15}$. Not surprisingly, $W(n,T)$ is inversely proportional to $H(n, T)$,
\begin{equation}
    W(n,T) \sim 
    \left\{\begin{array}{lll}
        T^{\frac{2}{n+2}}, & \,\,\,\,\,\,\,\,  (T\ll T_{cr}); \\ 
        &&    \\
        T^{\frac{1}{n}}, & \,\,\,\,\,\,\,\, (T\gg  T_{cr}). 
     \end{array}\right.       \label{width}
\end{equation}
This is expected from the conservation of area of $\rho^{(1)}(z)$ curve, $\int dz \rho^{(1)}(z) = 1$. 
Note that the crossover temperature $T_{cr} \sim 0.1$ is smaller than the value of Eq.(\ref{pow1}), reflecting that the shape of the curve $\rho^{(1)}(z)$ evolves with $n$.

The two-segment behavior of $H(n,T)$ and $W(n,T)$ as functions of $T$ shows that the oscillators' movement in KG model is dominated by different factors in the high and low temperature limits. In the high temperature limit, due to the large amplitude of oscillations, the local potential $|z|^{n}$ dominates the movement and the particles are more independent. The approximate energy conservation of a local oscillator gives $W^{n} \sim T/2$, leading to $W \sim T^{1/n}$. In the low temperature limit, the oscillators move near the bottom of the potential well and are influenced jointly by the local potential and the non-local interaction $(z_1-z_2)^2$ in a more subtle way, giving rise to the nontrivial behavior $W \sim T^{2/(n+2)}$.

From the two-segment power law behavior of the height $H(n,T)$ and width $W(n,T)$, one expects certain scaling behaviors in the function of $\rho^{(1)}(z, T)$ in the low and high temperature limits. In Fig.4(a), we plot $-\ln[\rho^{(1)}(z,T)/\rho^{(1)}(0,T)]$, the minus log of $\rho^{(1)}(z,T)$ rescaled by its height at $z=0$, as functions of $z$ for $n=6$ at temperatures from $10^{-6}$ to $10^{6}$. All curves behaves as $z^2$ for small $z$ and $z^6$ for large $z$, separated by a temperature dependent crossover $z_{cr}$. As temperatures increases, the curves move towards right at different speed in the low and high temperature regimes. 

In Fig.4(b) and (c), we separately plot the low and high temperature data of Fig.4(a), with $z$ rescaled by the width of $\rho^{(1)}(z)$, i.e., by $T^{2/(n+2)}$ in the low temperature limit, and by $T^{1/n}$ in the high temperature limit, respectively (see Eq.(\ref{width})). In Fig.4(b), all curves in the low temperatures $T=10^{-6} \sim T=10^{-1}$ nicely merge into a quadratic function in the small $z$ regime, once $z$ is scaled by the low temperature width power $T^{2/(n+2)}$. In the large $z$ regime, they do not merge into a power law curve. The crossover occurs at $z/T^{\frac{2}{n+1}} \sim 1$. We have checked that the above behavior holds for all $n>2$. The findings of Fig.4(b) can be summarized as the following scaling form for $n>2$,
\begin{eqnarray}   \label{rholowT}
\rho^{(1)}(z, T) =&& \rho_0  T^{-\frac{2}{n+2}} e^{- a \left[ z/T^{\frac{2}{n+2}} \right]^2},   \nonumber \\
   && \,\,\,\,\,\,\,\, \,\,\,\,  \,\,\,\,\,\,\,\, \,\,\,\,  \,\,\,\,  \,\,  (z/T^{\frac{2}{n+2}} \ll 1, \,\, T \ll 1).   
\end{eqnarray}
The coefficient $a \approx 1.0$ has a very weak dependence on $n$. 

In the high temperature limit, we scale $z$ by the high temperature width power $T^{1/n}$ and re-plot the high temperature curves of Fig.4(a) in Fig.4(c). We find that $\rho^{(1)}(z, T)$ has a scaling form only in the large $z$ regime. Considering that the non-local coupling plays no role in this limit, one can estimate $\rho^{(1)}(z_{i})$ by $\text{Tr}_{(i-1,i+1)} e^{-\beta \left[ |z_{i}|^{n}/n + (z_i- z_{i-1})^{2}/2 + (z_{i+1} - z_i)^2/2 \right] }$, giving $\rho^{(1)}(z_{i}) \sim e^{-\beta (|z_{i}|^{n}/n + z_i^{2})}$. This formula (dashed lines) nicely match the whole curves in Fig.4(c). We thus obtain the high temperature asymptotic behavior of $\rho^{(1)}(z, T)$ (for $n>2$) as
\begin{equation}   \label{rhohighT}
   \rho^{(1)}(z, T) = \rho_0  T^{-\frac{1}{n}} e^{- \frac{1}{T} \left( \frac{|z|^n}{n} + z^{2} \right)},  \,\,\,\,\, \,\,\,\, (T \gg 1).
\end{equation}
A more systematic derivation of Eq.(\ref{rhohighT}) from CVM equations is given in Appendix A. In the large $z$ limit, the $z^{n}$ term dominates the exponent (for $n  > 2$) and the above equation becomes a scaling form $\rho^{(1)}(z, T) \sim T^{-\frac{1}{n}} \rho^{(1)}(z/T^{\frac{1}{n}}, 1)$. This then gives a well-defined width $W(n,T) \sim T^{\frac{1}{n}}$, being consistent with Eq.(\ref{width}).

For $n=2$, the local and non-local potential have the same power. The above analysis for high temperature does not hold any more because the non-local interaction cannot be ignored even in the high temperature limit. However, the form of Eq.(\ref{rholowT}) covers Eq.(\ref{rhohighT}) in this case. So we have for $n=2$,
\begin{equation}   \label{rhon2}
   \rho^{(1)}(z, T)(n=2) = \rho_0  T^{-\frac{1}{2}} e^{- a \frac{z^2}{T}},
\end{equation}
Fitting the numerical data of $\rho^{(1)}(z,T)$ ($n=2$) with this equation gives $a \approx 1.1$.

\begin{figure}
 \vspace*{-1.0cm}
\begin{center}
  \includegraphics[width=370pt, height=280pt, angle=0]{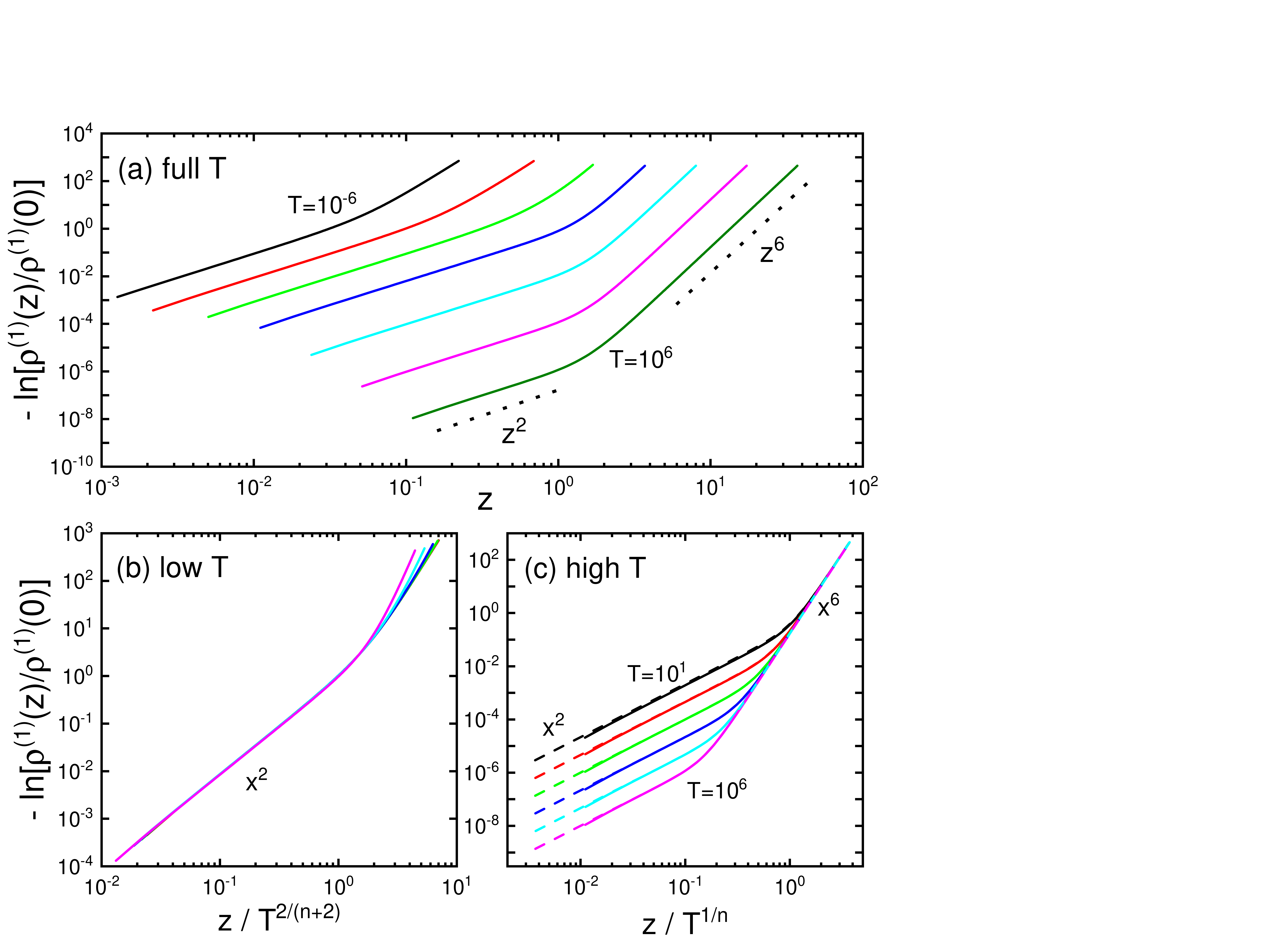}
  \vspace*{-1.0cm}    \label{Fig4}
\end{center} 
\caption{(color online) Curves of $-\ln[\rho^{(1)}(z)/\rho^{(1)}(0)]$ for $n=6$ as functions of rescaled $z$. (a) $-\ln[\rho^{(1)}(z)/\rho^{(1)}(0)]$ as functions of $z$ for $T= 10^{-6}$, $10^{-4}$, ... , $10^{6}$ (from left to right) Black dotted lines marked out the power law behavior. (b) $-\ln[\rho^{(1)}(z)/\rho^{(1)}(0)]$ as functions of $z/T^{2/(n-2)}$ for low temperatures $T=10^{-6} \sim 10^{-1}$. (c) $-\ln[\rho^{(1)}(z)/\rho^{(1)}(0)]$ as functions of $z/T^{1/n}$ for high temperatures $T=10^{1} \sim 10^{6}$ (from left to right). The dashed lines are curves for $-\beta(|z|^n/n + z^2)$ vs $z/T^{1/n}$. The power functions $x^{2}$ and $x^{6}$ are marked in the figures.
} 
\end{figure}

 \subsection{ $\langle |z|^m\rangle$ and $\langle (z_1-z_2)^2\rangle$ }
 
In this part, we study the thermodynamic averages $\langle |z|^m\rangle$ for arbitrary $m$ and the average of nearest-neighbor coupling $\langle (z_1-z_2)^2\rangle$.
$\langle |z|^{m} \rangle$ can be calculated directly from the single-site reduced density $\rho^{(1)}(z,T)$ through
 \begin{equation}   \label{avezm1}
     \langle |z|^m \rangle =  \int_{-\infty}^{+\infty} |z|^m  \rho^{(1)}(z, T) dz.
 \end{equation}
Using the low and high temperature expressions for $\rho^{(1)}(z,T)$ Eqs.(\ref{rholowT}) and (\ref{rhohighT}), we obtain the low and high temperature asymptotic behavior of $\langle |z|^{m} \rangle$ as
\begin{equation}
    \langle |z|^{m} \rangle \sim 
    \left\{
       \begin{array}{lll}
          T^{\frac{2m}{n+2}},  & \,\,\,\,\,\,\,\,  (T \ll T_{cr}); \\ 
          \\
          T^{\frac{m}{n}} ,  & \,\,\,\,\,\,\,\,   (T \gg T_{cr}). 
     \end{array} \right.       \label{avezm2}
\end{equation}
Note that in the high temperature limit, the width of $\rho^{(1)}(z,T)$ becomes so large that the integration in Eq.(\ref{avezm1}) will be dominated by the $|z|^{n}$ term in the expression of $\rho^{(1)}(z,T)$, Eq.(\ref{rhohighT}). So the high temperature behavior $\langle |z|^{m} \rangle \sim T^{m/n}$ is the properties of the atomic limit, as analyzed in Appendix A.

\begin{figure}
 \vspace*{-1.0cm}
\begin{center}   
  \includegraphics[width=300pt, height=250pt, angle=0]{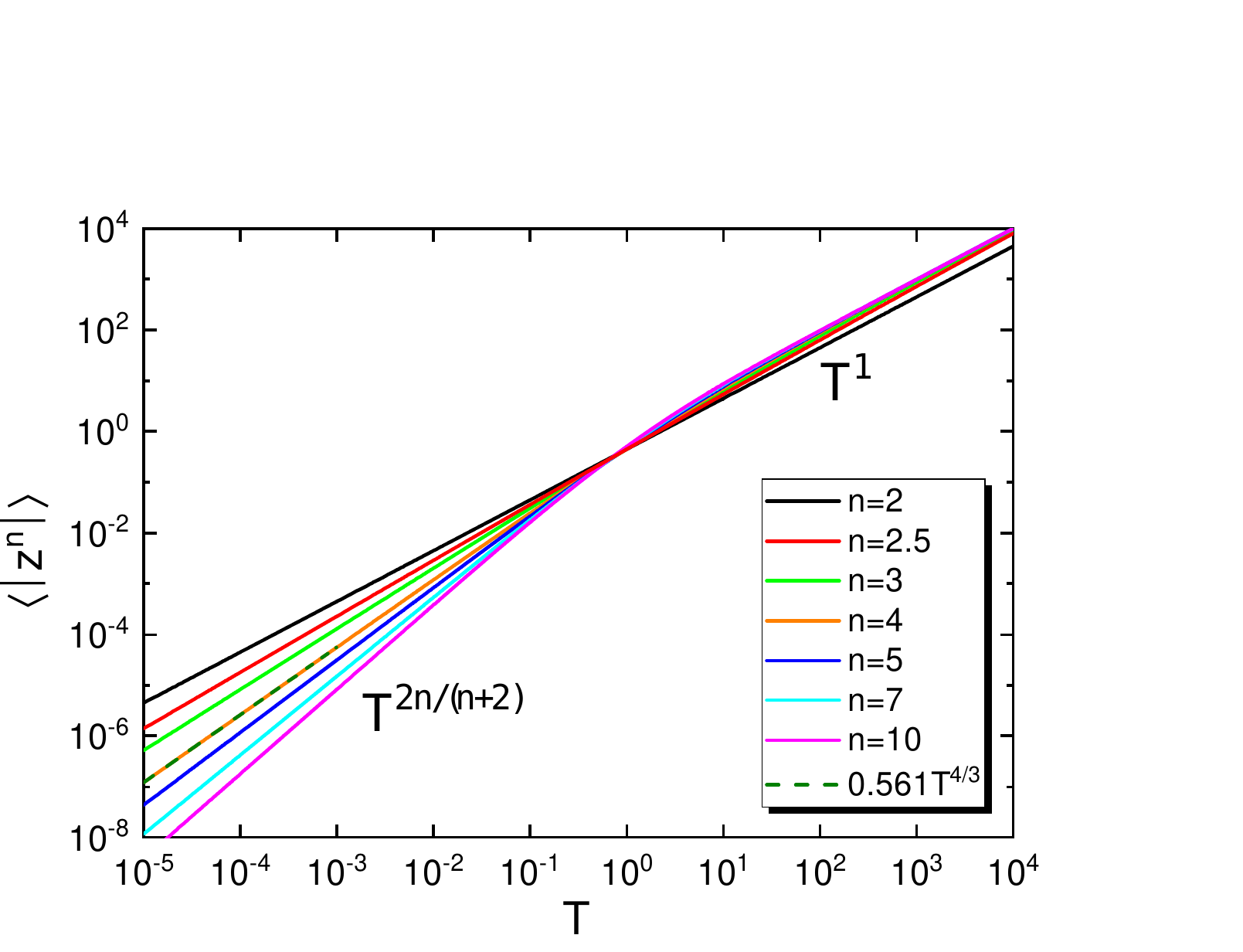}
  \vspace*{-1.0cm}
\end{center} 
\caption{(color online) $\langle |z^n|\rangle$ as functions of temperature $T$ for various $n$ values. The low and high temperature asymptotic power laws are marked in the figure.
The dashed line is for $0.561T^{4/3}$ obtained in Ref.\cite{Boyanovsky1} for $\phi^4$ lattice model (KG model at $n=4$) in the low temperature limit.
} \label{Fig5}
\end{figure}

\begin{figure}
 \vspace*{-1.0cm}
\begin{center}
  \includegraphics[width=290pt, height=230pt, angle=0]{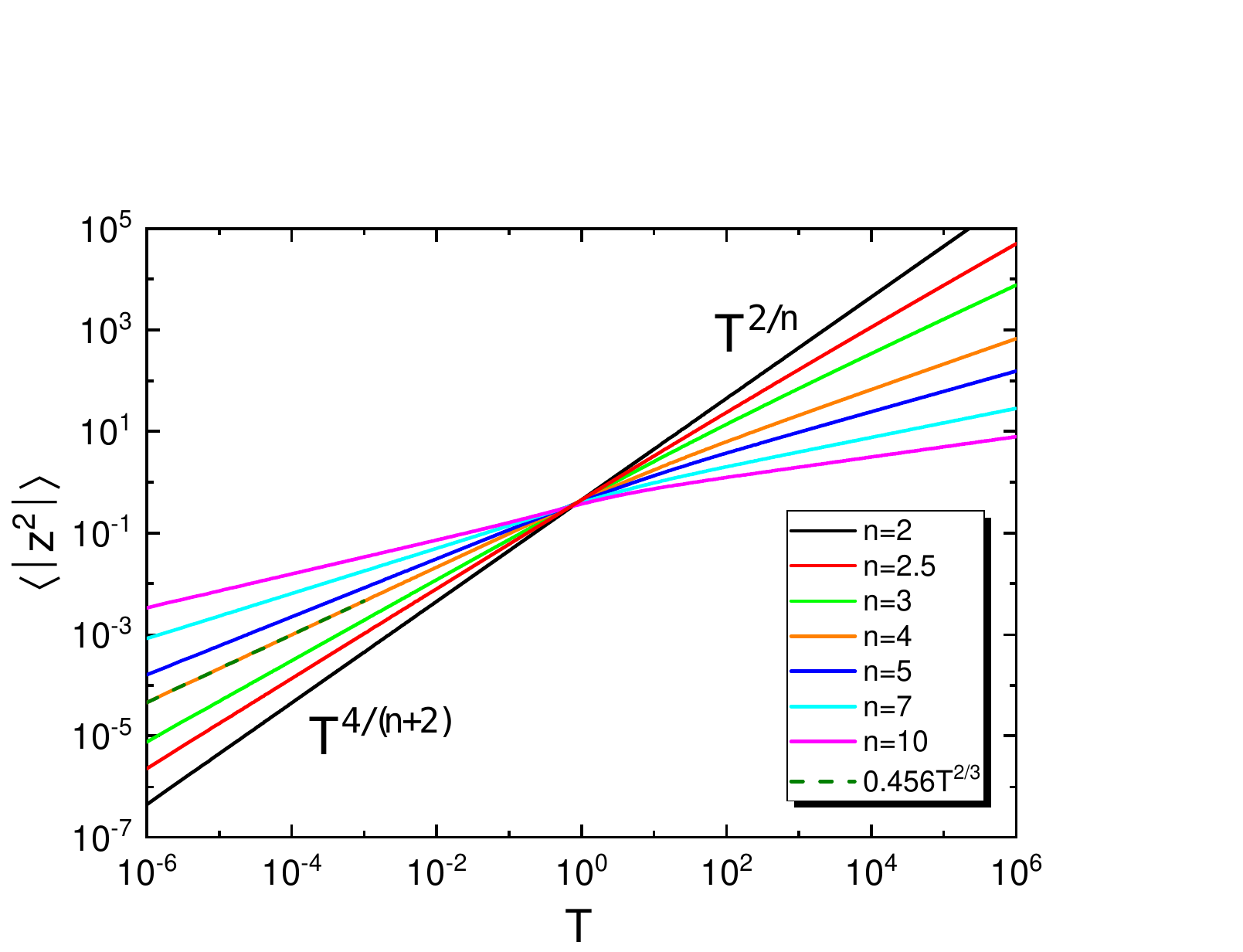}
  \vspace*{-1.0cm}
\end{center} 
\caption{(color online) $\langle |z^2|\rangle$ as functions of temperature $T$ for various $n$ values. The low and high temperature asymptotic power laws are marked in the figure.
The dashed line is for $0.456T^{2/3}$ obtained in Ref.\cite{Boyanovsky1} for $\phi^4$ lattice model ($n=4$) in the low temperature limit.
} \label{Fig6}
\end{figure}

In Fig.5 and Fig.6, we plot the numerical results for $\langle |z|^{n} \rangle(T)$ and $\langle |z|^{2} \rangle(T)$ respectively, for various $n$ values. These curves fulfil the asymptotic low and high temperature power laws in Eq.(\ref{avezm2}), with a crossover at around $T_{cr} \sim 1$. For $n=2$, the low and high temperature limits have same asymptotic power $T^{m/2}$ for arbitrary $m$. We did not observe any signature of crossover between the low and high temperature regime (see the $n=2$ curves in Figs.(\ref{Fig5}) and (\ref{Fig6})). This supports that for $n=2$, there is a unified power law function in the full temperature regime, i.e., Eq.(\ref{rhon2}).  For $n=4$, the exact low temperature asymptotic expression has been obtained as $\langle z^2 \rangle = 0.456...T^{2/3}$ and $\langle z^4 \rangle = 0.561...T^{4/3}$ \cite{Boyanovsky1}. As shown by the dashed lines in Fig.\ref{Fig5} and Fig.\ref{Fig6}, CVM data agree well with these exact formulas in the low temperature limit.

%
%

\begin{figure}
 \vspace*{-1.0cm}
\begin{center}
  \includegraphics[width=300pt, height=240pt, angle=0]{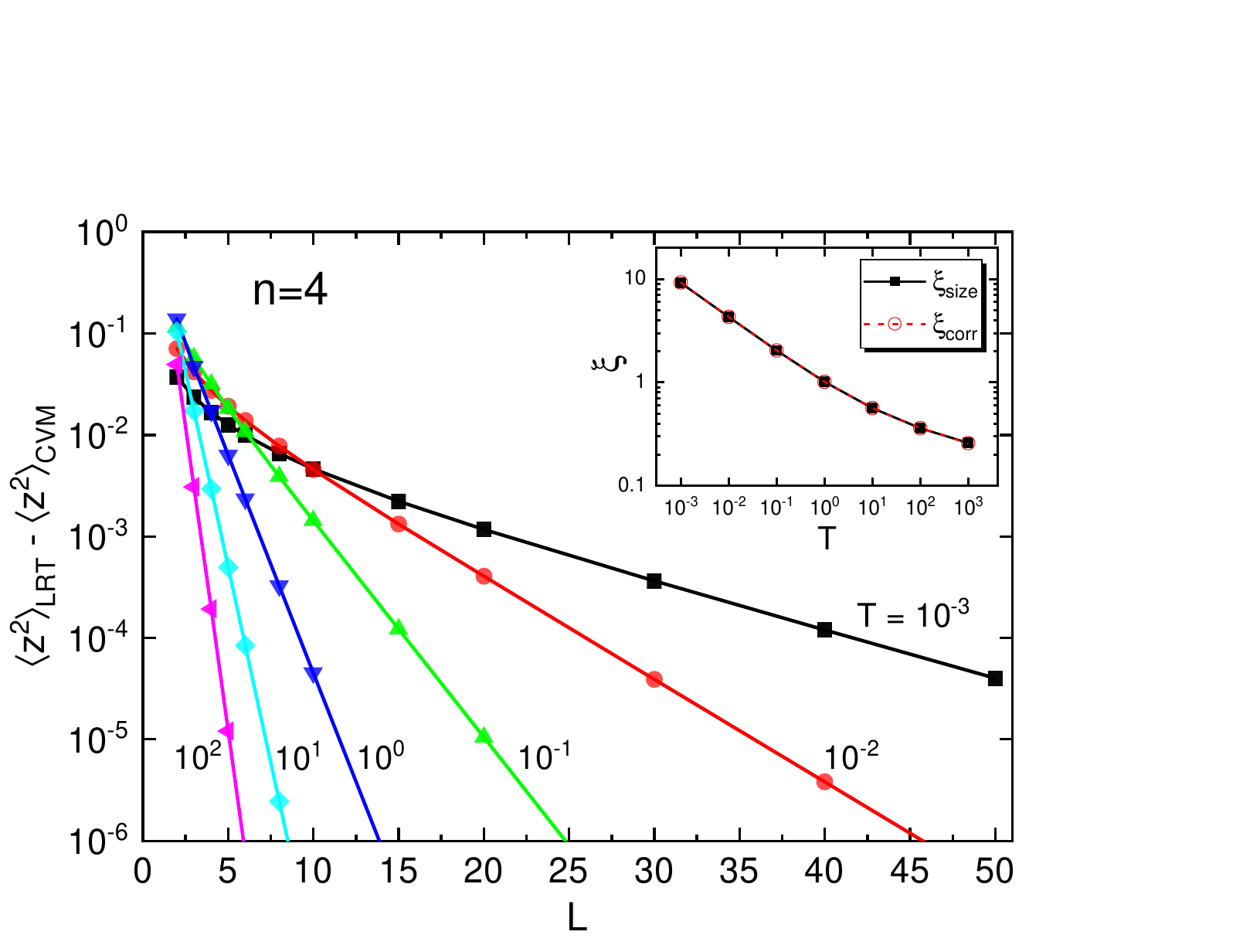}
  \vspace*{-1.0cm}
\end{center} 
\caption{(color online) The difference between $\langle z^{2} \rangle$'s obtained from LRT (Eq.(\ref{c1i_cal})) and CVM (Eq.(\ref{avezm1}) as functions of $L$ for $n=4$ and various temperatures (marked in  the figure). Inset: decay length $\xi_{\text{size}}(T)$ from fitting the exponential decay in the main figure (squares with solid guiding line), and the correlation length $\xi_{\text{corr}}(T)$ from fitting the correlation function via Eq.(\ref{cor_fun_def}) (cycles with dashed guiding lines).
} \label{Fig7}
\end{figure}

Here, it is appropriate to discuss the self-consistency between CVM and LRT. As noted in the end of section III.A and III.B, the system size $L$ does not appear in our CVM formula for the periodic boundary condition. It appears in LRT formula only through the summation of $k$ in Eq.(\ref{c1i_cal}). The LRT result from Eq.(\ref{c1i_cal}) should be consistent with the CVM result from $\rho^{(1)}(z)$ (Eq.(\ref{rho1_a})) and $\rho^{(2)}(z,w)$ (Eq.(\ref{rho1_b})) in the limit $L = \infty$ where both are expected to be exact. This consistency is a test of exactness of the theory. We consider the local quantity $\langle z^{2} \rangle$ which can be calculated from CVM and LRT respectively through
\begin{eqnarray}
&&  \langle z^{2} \rangle_{\text{CVM}} =  \int_{-\infty}^{+\infty} z^2  \rho^{(1)}(z, T) \, dz,  \nonumber \\
&&   \langle z^{2} \rangle_{\text{LRT}} = C_{11} = \frac{1}{L} \sum_{k} C(k).
\end{eqnarray}
Here, $C_{11}$ and $C(k)$ are given in Eq.(\ref{c1i_def}) and (\ref{c1i_cal}), respectively. 
It turns out that for finite $L$, LRT always produce larger $\langle z^2 \rangle$ than CVM. 
We gauge the finite size error in $\langle z^2 \rangle_{\text{LRT}}$ by $\delta \langle z^{2} \rangle \equiv \langle z^2 \rangle_{\text{LRT}}- \langle z^2 \rangle_{\text{CVM}}$. In Fig.\ref{Fig7}, we plot $\delta \langle z^{2} \rangle$ as functions of $L$ for a series of temperatures at $n=4$. They decay exponentially with increasing $L$, tending to zero in the thermodynamical limit. This confirms the expected consistency between CVM and LRT. We define the decay length $\xi_{\text{size}}$ as $\delta \langle z^2 \rangle = c e^{-L/\xi_{\text{size}}}$. The inset of Fig.\ref{Fig7} shows that $\xi_{\text{size}}(T)$ decreases with increasing $T$ and coincides with the correlation length $\xi_{\text{corr}}(T)$ extracted from the correlation function $C_{1i} \propto e^{-(i-1)/\xi_{\text{corr}} }$ in the large $i$ limit. This is a reasonable result since the finite size error can be measured by how much long-range correlation beyond the system size is neglected in the finite size calculation. It is therefore controlled by $L/\xi_{\text{corr}}$.

It should be noted that $\langle z^2 \rangle_{\text{LRT}}$ is not the exact result for a chain with finite length $L$. This is because CVM is exact for system with an open boundary condition, while in the derivation of LRT, we have used the Fourier transformation which applies only to periodic systems. Thus the combination of CVM and LRT is exact only in the thermodynamical limit. In fact, we have checked that for small $L$, $\langle z^2 \rangle_{\text{LRT}}(T)$ has an incorrect power law in the temperature dependence in the low temperature limit.

%

\begin{figure}
 \vspace*{-1.5cm}
\begin{center}
  \includegraphics[width=300pt, height=240pt, angle=0]{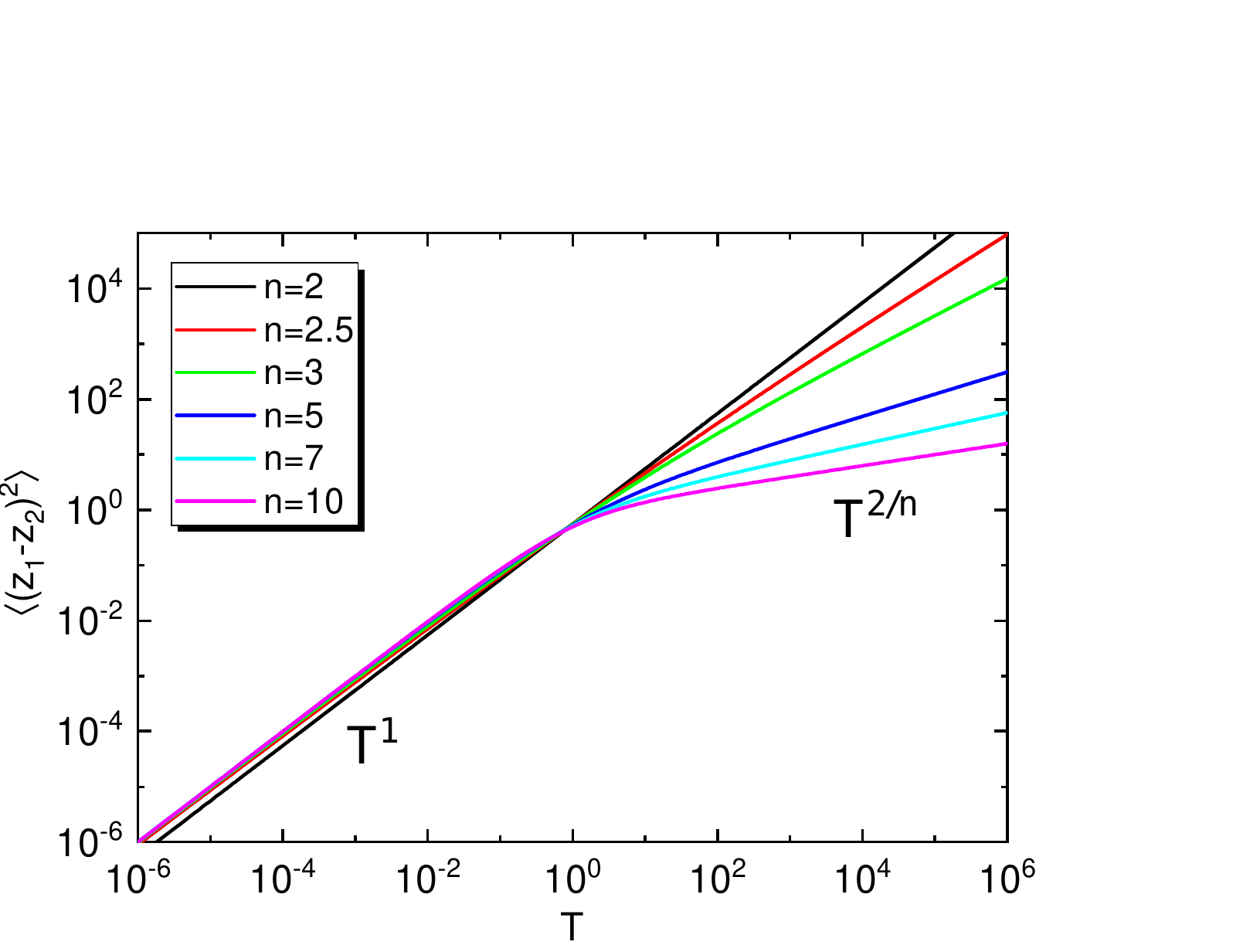}
  \vspace*{-1.0cm}
\end{center} 
\caption{(color online) $\langle (z_1-z_2)^2 \rangle$ as functions of temperature $T$ for various $n$ values. The low and high temperature asymptotic power laws are marked in the figure.
} \label{Fig8}
\end{figure}

The average $\langle (z_1-z_2)^2\rangle$ involves the correlation between the displacements $z_1$ and $z_2$ of the two nearest sites. It must be calculated from $\rho^{(2)}(z,w)$ of Eq.(\ref{rho1_b}). In Fig.8, $\langle (z_1-z_2)^2\rangle(T)$ as functions of $T$ is plotted for various $n$ values. Similar two-segment behavior is observed, with the asymptotic behavior extracted as
\begin{equation}         \label{avez1z2}
    \langle (z_1-z_2)^2\rangle \sim 
    \left\{
       \begin{array}{lll}
          T^{1},  & \,\,\,\,\,\,\,\,   (T \ll T_{cr}); \\ 
          \\
          T^{\frac{2}{n}} ,  & \,\,\,\,\,\,\,\,  (T \gg T_{cr}),
     \end{array} \right. 
\end{equation}
with $T_{cr} \sim 1.0$.
In the high temperature limit, the correlations between neighboring sites can be neglected, we have $\langle (z_1-z_2)^2 \rangle \sim 2 \langle z^2 \rangle \sim T^{2/n} $. In the low temperature limit,  $\langle (z_1-z_2)^2 \rangle = 2 (\langle z^2 \rangle - \langle z_1 z_2 \rangle)$.  Since for $n>2$ and $T \rightarrow 0$, $\langle z^2 \rangle \sim T^{4/(n+2)}$ is much larger than $\langle (z_1 -z_2)^2 \rangle \sim T^{1}$, we expect that the $T^{1}$ behavior in the latter is the result of a cancellation of the leading $T^{4/(n+2)}$ term in both $\langle z^2 \rangle$ and $\langle z_1 z_2 \rangle$. This shows that in the low temperature limit, the nearest-neighbor correlation of $x_i$ is as large as the local correlation. We see that the temperature behavior of the averages is characterized by a two-segment fashion, with the high $T$ properties dominated by the local potential, and the low $T$ properties determined jointly by the local potential and the non-local interaction.

\subsection{Specific Heat}

The internal energy $\langle H \rangle$ for the KG lattice model is expressed in terms of the averages discussed in previous subsection,
\begin{equation}   \label{aveH}
    \langle H \rangle =\frac{1}{2}TL + \frac{L}{2} \langle(z_1-z_2)^2 \rangle +\frac{L}{n} \langle |z|^n \rangle.
\end{equation}
The specific heat can be obtained from $C_v = (1/L) \partial \langle H \rangle / \partial T$.
%
%
\begin{figure}
 \vspace*{-1.8cm}
\begin{center}
  \includegraphics[width=350pt, height=240pt, angle=0]{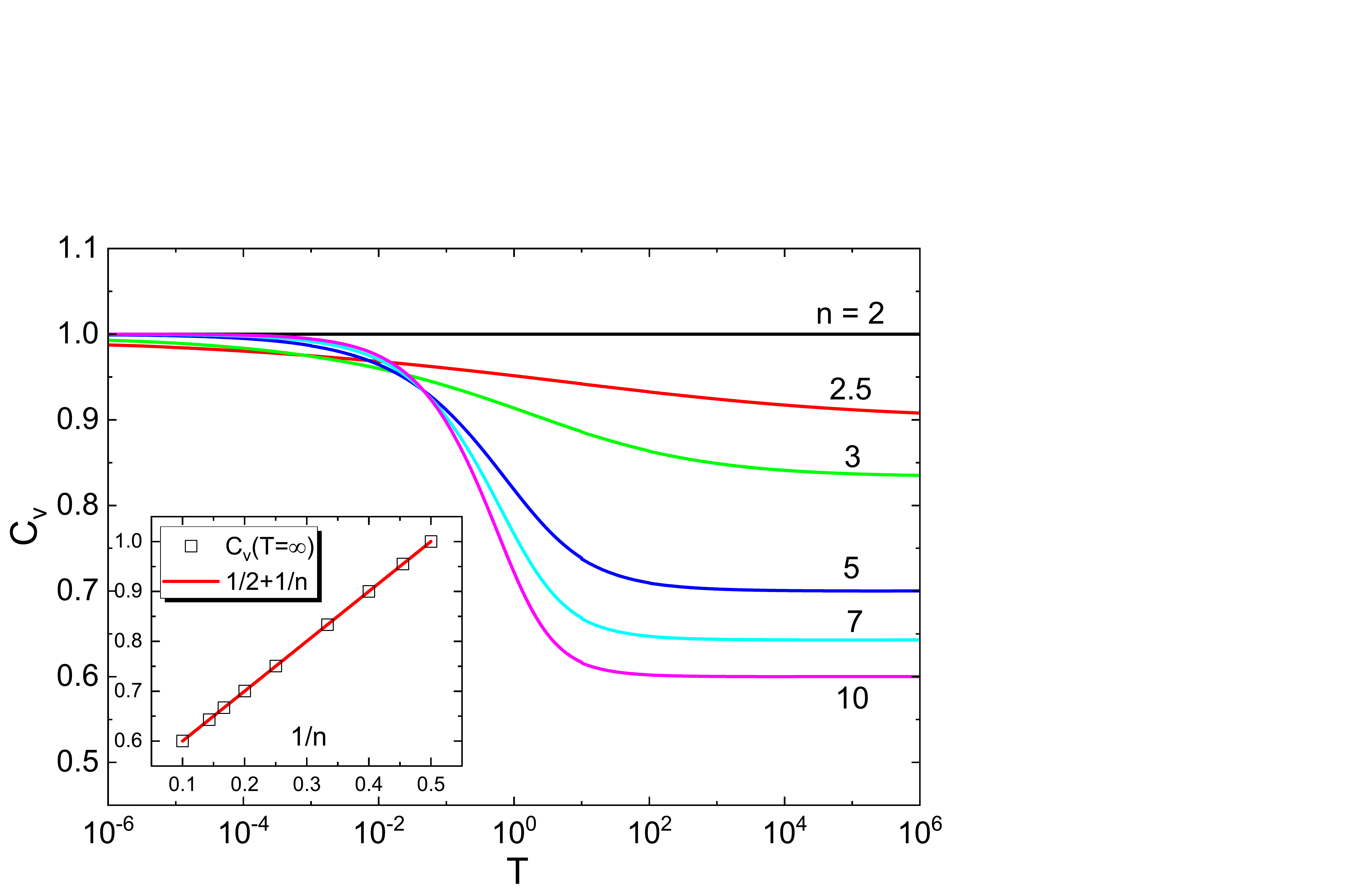}
  \vspace*{-1.0cm}
\end{center} 
\caption{(color online) Specific heat $C_v$ as functions of temperature $T$ for various $n$ values. The inset shows the extrapolated value of $C_v(\infty)$ (squares) versus $1/n$ (solid line), showing $C_v(\infty) = 1/2 + 1/n$. 
} \label{Fig9}
\end{figure}
%
In Fig.\ref{Fig9}, we plot $C_v$ as functions of $T$ for various $n$ values. For $n>2$, $C_v(T)$ is a decreasing function of $T$. We observed $C_v=1.0$ at $T=0$ and it decreases to a constant smaller than unity in the high temperature limit. The crossover temperature separating the low and the high temperature plateau is about $T_{cr} = 10^{-2} \sim 10^{0}$ for $2 < n < 10$. For the special case $n=2$, $C_v = 1$ for all temperatures, as expected from the energy equipartition theorem. The inset shows that $C_v(T=\infty)$ agrees well with $1/2 + 1/n$. For $n=4$, the curve $C_v(T)$ and the high temperature value $3/4$ agree with the results from projective truncation approximation for $\phi^4$ lattice model \cite{PTAphi4}. The obtained low and high temperature asymptotic values of $C_v$ for $n>2$ are thus summarized as
\begin{equation}
     C_v=\left\{
     \begin{array}{lll} 
        1, & \,\,\,\,\,\,\,\, (T  \ll 1); \\
        && \\
        \frac{1}{2}+\frac{1}{n}, & \,\,\,\,\,\,\,\,   (T\gg 1).
     \end{array}\right.     \label{powB}
 \end{equation}

This result can be understood in terms of previous results. According to Eqs.(\ref{avezm2}) and (\ref{avez1z2}), we write the asymptotic expression for $\langle (z_1 - z_2)^2 \rangle$ and $\langle |z|^{n} \rangle$ as
\begin{equation}
    \langle (z_1-z_2)^2\rangle \sim 
    \left\{
       \begin{array}{lll}
          A \, T^{1},  & \,\,\,\,\,\,\,\, (T \ll 1);  \\ 
          \\
          \tilde{A} \, T^{\frac{2}{n}} ,  & \,\,\,\,\,\,\,\, (T \gg 1). 
     \end{array} \right.       \label{avez1z2coef}
\end{equation}
and
\begin{equation}
    \langle |z|^{n} \rangle \sim 
    \left\{
       \begin{array}{lll}
          B \, T^{\frac{2n}{n+2}},  & \,\,\,\,\,\,\,\, (T \ll 1); \\ 
          \\
          \tilde{B} \, T^{1} ,  & \,\,\,\,\,\,\,\, (T \gg 1). 
     \end{array} \right.       \label{avezncoef}
\end{equation}
Here, $A$, $\tilde{A}$, $B$, and $\tilde{B}$ are temperature-independent coefficients in the asymptotic limits.
Inserting the above two equations into Eq.(\ref{aveH}), we obtain for $n > 2$
\begin{equation}    \label{C_v1}
C_v=\left\{
 \begin{array}{lll}
   \frac{1}{2} + \frac{1}{2}A + \frac{2}{n+2}B \, T^{\frac{n-2}{n+2}},   & \,\,\,\,\,\,\,\,  (T \ll 1);    \\
   & \\
\frac{1}{2}+ \frac{1}{n} \tilde{B} +\frac{1}{n} \tilde{A} \, T^{\frac{2}{n}-1} ,   & \,\,\,\,\,\,\,\, (T \gg 1).
\end{array}
\right.   
\end{equation}
For $T \ll 1.0$, the systems becomes quadratic and equipartition of energy gives $A=1.0$. For $T \gg 1.0$, from the high temperature asymptotic expression of $\rho^{(1)}(z,T)$ in Eq.(\ref{rhohighT}), one can rigorously derive $\tilde{B}=1$. Inserting these into Eq.(\ref{C_v1}), one obtains the observed result Eq.(\ref{powB}) for $n>2$. For $n=2$, we expect that $A(n=2)+B(n=2)=1$ and $\tilde{A}(n=2) +\tilde{B}(n=2) = 1$, as required by $C_v(n=2) = 1.0$.  The crossover temperature of $C_v(T)$ can be estimated by equating the low and high temperature expressions in Eq.(\ref{C_v1}).

\subsection{Static Correlation Function}

\begin{figure}
	\vspace*{-1.5cm}
	\begin{center}
		\includegraphics[width=380pt, height=300pt, angle=0]{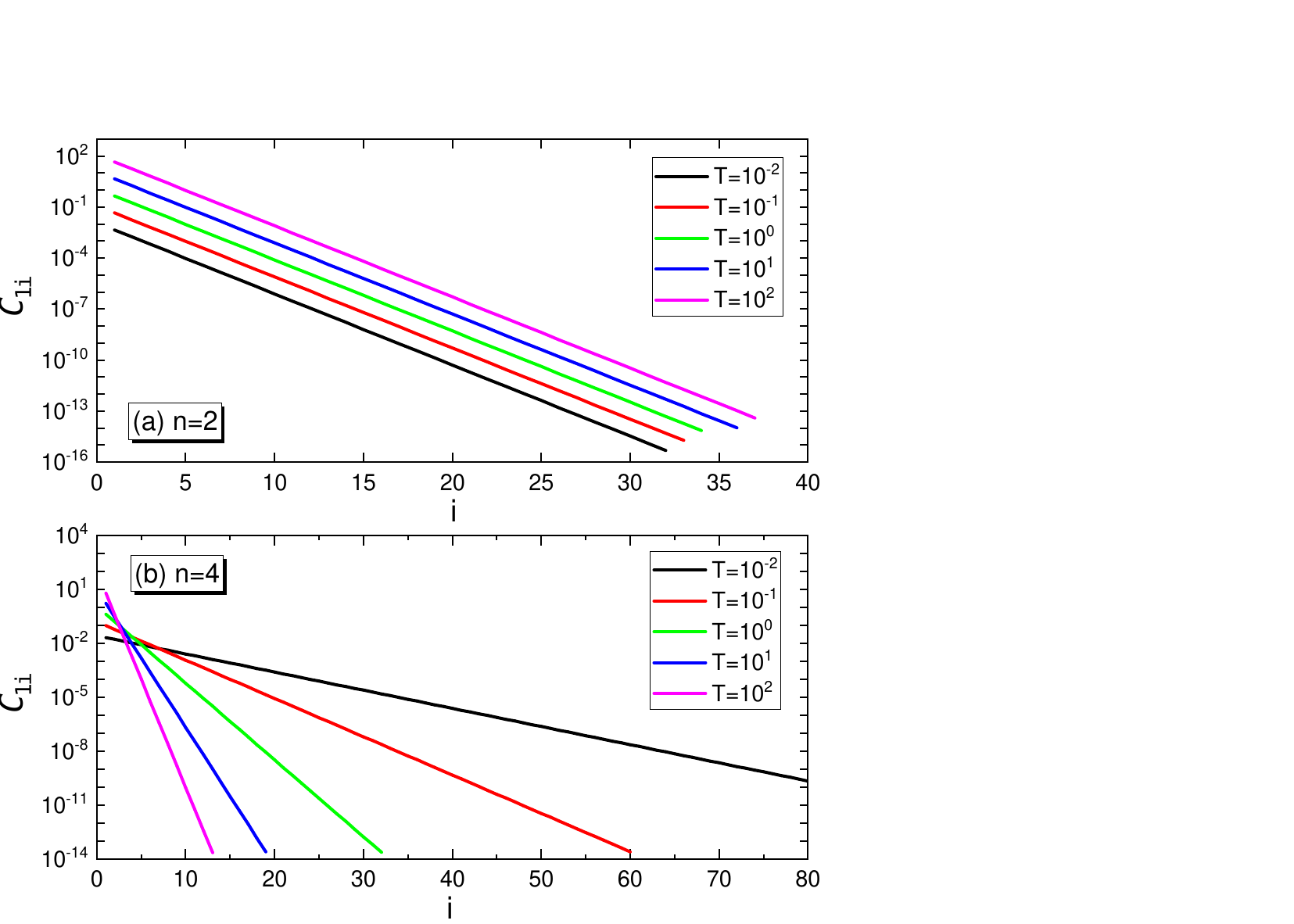}
		\vspace*{-1.0cm}
	\end{center} \caption{(color online) Correlation function $C_{1i}$ as functions of lattice site index $i$, for (a) $n=2$ and (b) $n=4$ at various temperatures. 
	} \label{Fig10}
\end{figure}

The static correlation function $C_{1i} = \langle x_1 x_i \rangle - \langle x_1 \rangle \langle x_i \rangle$ is calculated from LRT formula Eqs.(\ref{ck}) and (\ref{c1i_cal}) for a sufficiently long chain, choosing the function $B(x) \equiv x$. For the present model, $\langle x_i \rangle = 0$. Fig.\ref{Fig10} shows that $C_{1i}$ decays exponentially with increasing $i$ for any temperature, starting from $i=1$ and until $i$ is close to $L/2$. We therefore define the correlation length $\xi$ as
\begin{equation}      \label{cor_fun_def}
   C_{1i} = \langle x^2\rangle  e^{-\frac{i-1}{\xi}}, \,\,\,\,\,\,\,\, (1 \leq i \ll L/2).
\end{equation}
Fig.\ref{Fig10}(a) shows that for $n=2$, $\xi$ is a constant independent of $T$. Fig.\ref{Fig10}(b) shows that for $n=4$, $\xi$ is a decreasing function of temperature. 

\begin{figure}
 \vspace*{-2.0cm}
\begin{center}
  \includegraphics[width=300pt, height=240pt, angle=0]{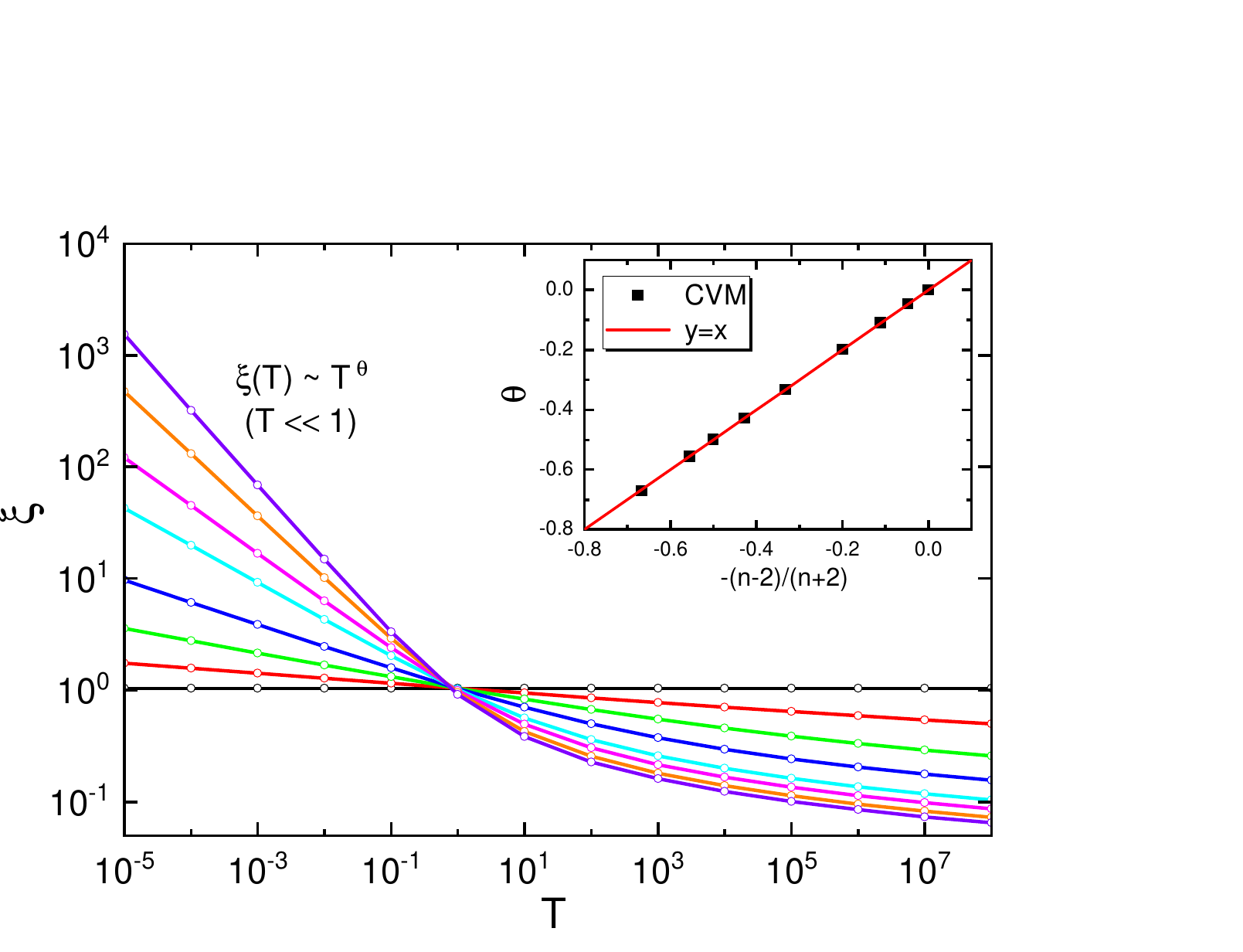}
  \vspace*{-1.0cm}
\end{center} \caption{(color online) Correlation length $\xi$ as functions of temperature $T$ for various $n$ values. From bottom to top on  the left side, $n= 2.0$, $2.2$, $2.5$, $3.0$, $4.0$, $5.0$, $7.0$, and $10.0$. The symbols are data and lines are for guiding eyes. Inset: low temperature asymptotic power $\theta$ from $\xi(T \ll T_{cr}) \sim T^{\theta}$ versus $-(n-2)/(n+2)$ (solid squares), compared with the function $y=x$ (red line).  
} \label{Fig11}
\end{figure}

The temperature dependence of correlation length $\xi$ is presented in Fig.\ref{Fig11} for various $n$ values. For $n=2$, our fitting gives the $T$-independent value $\xi(T) = 1.03904$, in good agreement with the exact value $1/\ln{[(3+\sqrt{5})/2]}$ (see the PTA result in Sec.IV.E). For $n> 2$, $\xi(T)$ diverges in a power law fashion $\xi \sim T^{\theta < 0}$ in the low temperature limit. In the large $T$ limit, $\xi(T)$ decreases slower than power law. It is interesting to notice that the curves of $\xi(T)$ for different $n$ values cross a common point at $T=T_c \approx 1.0$ and $\xi_c \approx 1.0$.

\begin{figure}
 \vspace*{-2.5cm}
\begin{center}
  \includegraphics[width=310pt, height=260pt, angle=0]{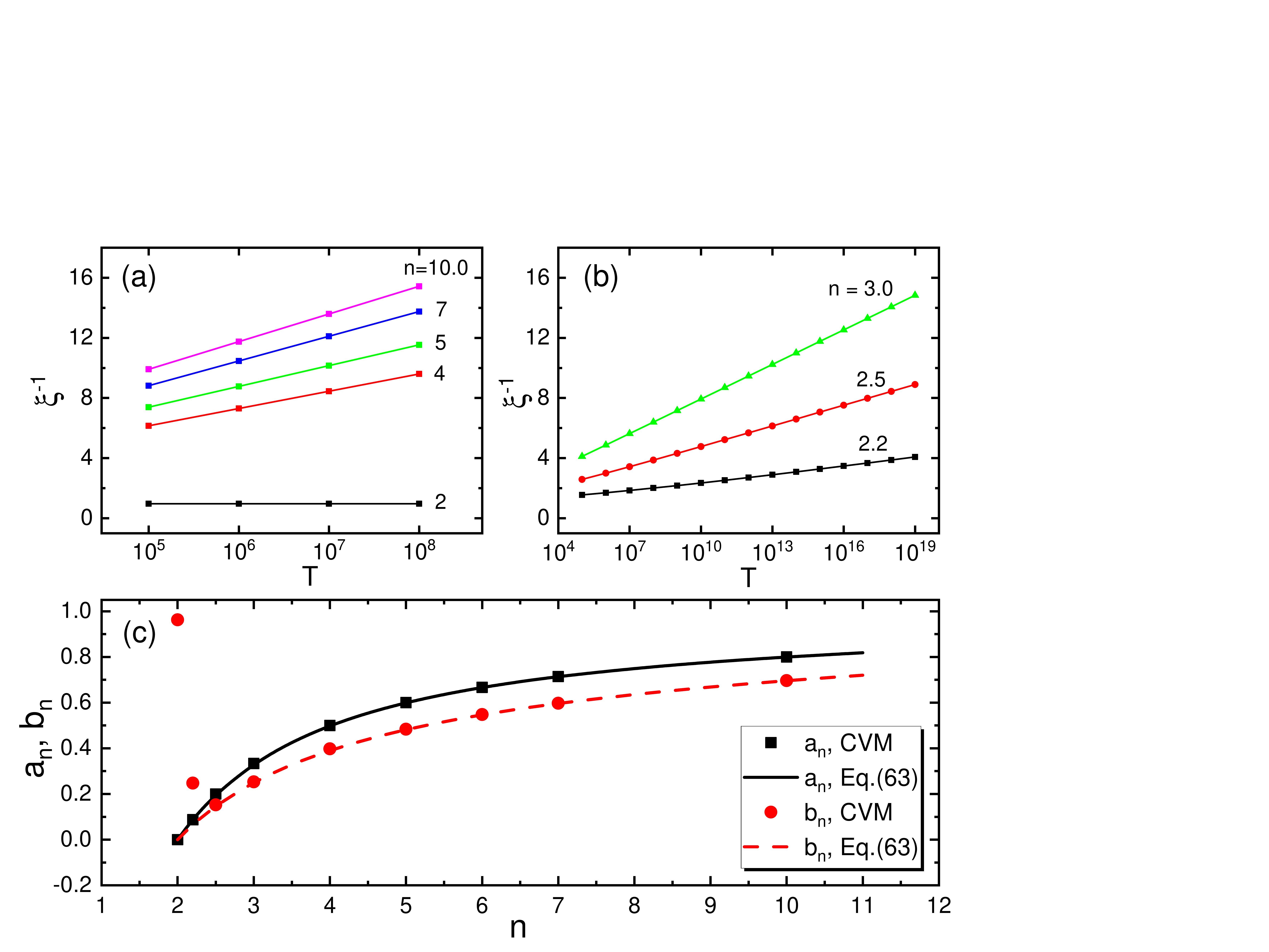}
  \vspace*{-1.0cm}
\end{center} \caption{(color online) (a) and (b): Inverse correlation length $\xi^{-1}$ as functions of $T$ in the high temperature regime for various $n$ values. The $n$ values are marked in the figure. In (c), the fitted power $a_n$ and $b_n$ from CVM data according to $\xi^{-1}(T \gg T_{cr}) = a_n \ln{T} + b_n$ (squares and dots) are compared to formula Eq.(\ref{pow_h2}) (solid and dashed lines).
} \label{Fig12}
\end{figure}

Numerical fitting reveals the following asymptotic power law behavior of $\xi(T)$ in the low temperature limit,
\begin{equation}   \label{pow_low}
    \xi(T) \sim  T^{-\frac{n-2}{n+2}}, \,\,\,\,\,\,\,\,  (T\ll T_{cr}).
\end{equation}
See inset of Fig.\ref{Fig11}. In the high temperature regime, combining the numerical fitting with the asymptotic analysis in Appendix A, we extract the asymptotic behavior of $\xi(T \gg T_{cr})$ (see Fig.\ref{Fig12}) as
\begin{equation}     \label{pow_h}
     \xi(T)= \frac{1}{a_n \ln T +b_n}, \,\,\, \,\,\,\,\,\,\, (n>2, \,\, T \gg T_{cr}) ,
 \end{equation}
with
\begin{eqnarray}      \label{pow_h2}
a_n &=& \frac{n-2}{n},    \nonumber \\
b_n &=&  \ln\left[ n^{-\frac{2}{n}}\frac{\Gamma(\frac{1}{n})}{\Gamma(\frac{3}{n})}\right].
\end{eqnarray}
In Fig.\ref{Fig12}(a) and (b), we show the high temperature data for $1/\xi(T)$ for various $n$ values, showing that they are linear function of $\ln{T}$ for sufficiently large $T$. The smaller $n$ is, the larger $T$ is needed to observed the linearity, as seen by comparing data in Fig.\ref{Fig12}(a) and (b).
In Fig.\ref{Fig12}(c), the numerical fitted coefficient $a_n$ and $b_n$ agree well with Eq.(\ref{pow_h2}) except for $n$ very close to $2$. This is because for $n \rightarrow 2+$, one needs to go to extremely high $T$ to observe the behavior of Eq.(\ref{pow_h}). In Appendix A, we give the details of the derivation of Eqs.(\ref{pow_h}) and (\ref{pow_h2}).

The crossover temperature $T_{cr}$ in $\xi(T)$ can be estimated by equating Eqs.(\ref{pow_low}) and (\ref{pow_h}). We obtain $T_{cr} = c^{n/(n-2)}$ with $c > 1$ being a constant. As $n$ decreases towards $n = 2^{+}$, $T_{cr}$ diverges and the low temperature formula Eq.(\ref{pow_low}) governs the whole temperature range. This is why for smaller $n$, one needs to go to larger $T$ to observe the high temperature asymptotic behavior Eq.(\ref{pow_h2}). For $n=2$, $T_{cr} = \infty$. Eq.(\ref{pow_low}) gives out $T$-independent $\xi$, being consistent with the $n=2$ curve in Fig.(\ref{Fig11}).

\subsection{Comparison with PTA }

In this section, we present and analyze the PTA results for the one-dimensional KG lattice model. As was discussed in detail in Ref.\cite{PTAphi4} for the one-dimensional $\phi^{4}$ lattice model (corresponding to KG model with $n=4$), PTA can be used to systematically study classical many-body problems. The PTA based on the simplest operator basis $B_1 = \{ Q_{k} = 1/\sqrt{L} \sum_{j=1}^{L} x_j e^{-ijk}\}$ (denoted as PTA-$B_1$ below) is shown to be equivalent to the quadratic variational method. This method has only been tested at certain points of parameters. 
Here, using the exact CVM results for KG model as a reference, we analyze the accuracy of PTA in the full temperature range. Demonstrating with $n=4$ case, we show that PTA-$B_1$ already gives qualitatively correct low and high temperature asymptotic behaviors. The PTA-$B_2$ based on basis $B_2 = \{ Q_{k}, R_{k} \}$, with $R_k = 1/\sqrt{L} \sum_{j=1}^{L} x^3_j e^{-ijk}$, further improves the quantitative accuracy. The analysis for PTA-$B_1$ equations also provides a derivation of the low temperature asymptotic exponents of the averages $\langle x_i^{m} \rangle(T)$ and the correlation length $\xi(T)$, as presented in Appendix B.

To carry out PTA, one first selects a set of operators (dynamical variables for the case of classical models) as the basis. The equation of motion of Green's functions defined on these operators are then truncated by using projection method. Here, we sketch the main formula for the one-dimensional KG model $H(K, \gamma)$ defined in Eq.(\ref{ham}) for even $n$. The final asymptotic expression also applies to the odd $n$ case. The general formalism of PTA for classical models can be found in Ref.\cite{PTAphi4}. 

The $B_1$ basis $\{ Q_{k} \}$ is defined as
\begin{equation}
   Q_{k} = \frac{1}{\sqrt{L}} \sum_{j=1}^{L} x_j e^{-ijk}.
\end{equation}
Here $k = (2\pi/L) m_k$ ($m_k = 0, 1, ..., L-1$) is the lattice momentum.
We used the inner product \cite{PTAphi4} $\left(X| Y \right) \equiv \langle \{ X^{\ast}, \{ Y, H \} \} \rangle$. Here $\{A,B\}$ is the Poisson bracket between variables $A$ and $B$. We obtain the inner product matrix ${\bf I}$ and Liouville matrix ${\bf L}$ as
\begin{eqnarray}
  {\bf I}_{k,k^{\prime}} &\equiv & \left(Q_k|Q_{k^{\prime}} \right) = \frac{1}{m} \delta_{k k^{\prime}},  \nonumber \\
  {\bf L}_{k,k^{\prime}} &\equiv& -\left(Q_k|\{ \{Q_{k^{\prime}}, H\} ,H \} \right)   \nonumber \\
 &=& \frac{1}{m^{2}} \left[ \omega_{0}^2(k) + (n-1)\gamma \langle x^{n-2} \rangle \right] \delta_{k k^{\prime}}.
\end{eqnarray}

The Green's function $G(Q_k|Q_{k}^{\ast})_{\omega}$ is then obtained approximately as
\begin{equation}
G(Q_k|Q_{k}^{\ast})_{\omega} \approx \frac{1/m}{\omega^2 - \frac{1}{m} \left[ \omega_{0}^2(k) + (n-1) \gamma \langle x^{n-2} \rangle \right]}.
\end{equation}
Here, $\omega^2_{0}(k) = 2K \left[ 1- \cos(k) \right]$.
The averages $\langle Q_k^{\ast}Q_k \rangle$ can be calculated from the spectral theorem. It reads
\begin{equation}     \label{Ck_pta}
\langle Q_k^{\ast}Q_k \rangle \approx \frac{1/ \beta}{\omega_{0}^2(k) + (n-1) \gamma \langle x^{n-2} \rangle}.
\end{equation}
This gives the correlation function
\begin{equation}  \label{corr-PTA}
\langle x_l x_j \rangle \approx \frac{1}{\beta L} \sum_{k} \frac{e^{ik(j-l)}}{\omega_{0}^2(k) + (n-1)\gamma \langle x^{n-2} \rangle}
\end{equation}
and the average $\langle x^2 \rangle$
\begin{equation}    \label{avex2pta}
\langle x^2 \rangle \approx \frac{1}{\beta L} \sum_{k} \frac{1}{\omega_{0}^2(k) + (n-1) \gamma \langle x^{n-2} \rangle}.
\end{equation}

To solve the unknown $\langle x^{n-2} \rangle$ appearing in the above equations self-consistently, we need to extend the EOM to the Green's functions $G\left( Q_k | R^{\ast}(m)_k \right)_{\omega}$, with $R(m)_k \equiv 1/\sqrt{L} \sum_j (x_j)^{m} e^{-ijk}$. We have $Q_k = R(1)_k$. The EOM and PTA for these extended GFs, together with the spectral theorem, lead to the following recursive relation for arbitrary odd $m > 0$,
\begin{equation}
   \langle x^{m+1} \rangle \approx m \langle x^2 \rangle \langle x^{m-1} \rangle.
\end{equation}
Using this equation, we can simplify Eq.(\ref{avex2pta}) into
\begin{equation} \label{x2-PTA}
\langle x^2 \rangle \approx \frac{1}{\beta L} \sum_{k} \frac{1}{\omega_{0}^2(k) + (n-1)!! \gamma \langle x^2 \rangle^{n/2-1}}.
\end{equation}

From the above equations for $\langle x^2 \rangle$ and $\langle Q_k^{\ast} Q_k\rangle$, we can analyze the low and high temperature asymptotic behavior of the averages $\langle x^m \rangle(T)$ and $\xi(T)$ from PTA-$B_1$. We put the details of analysis into Appendix B and only present the results here. 
For $n>2$, the asymptotic behavior of $\langle x^m \rangle$ (for even $m$) read
\begin{equation} 
    \langle x^m \rangle_{\text{PTA}}(T) \approx
    \left\{
       \begin{array}{lll}
          c_1 \, T^{\frac{2m}{n+2}},  & \,\,\,\,\,\,\,\, (T \ll 1);  \\ 
          \\
          c_2 \, T^{\frac{m}{n}} ,  & \,\,\,\,\,\,\,\, (T \gg 1). 
     \end{array} \right.       \label{avezm_pta}
\end{equation}
Here, $c_1 = (m-1)!! \left[16K\gamma (n-1)!! \right]^{-m/(n+2)}$, and
$c_2 = (m-1)!! \left[ (n-1)!! \gamma \right]^{-m/n}$.
Comparing them with the CVM expression Eq.(\ref{avezm2}), we find that PTA gives the correct asymptotic exponents.

The correlation length $\xi$ is independent of $T$ at $n=2$. It can be obtained exactly by comparing Eq.(\ref{Ck_pta}), which gives exact correlation function $C(k)$ for $n=2$, and the fitting expression $C^{\text{fit}}(k)$ in Eq.(\ref{Ck_fit}) in Appendix A. We obtain $\xi(n=2) = 1/\ln{ [(2+r + \sqrt{r^2 + 4r})/2]}$, with $r=\gamma/K$. For $K=\gamma=1$, it agrees well with the CVM numerical data $1.03904$ in Fig.\ref{Fig11}.

For $n \geq 2$, the asymptotic expression for the correlation length from PTA-$B_1$ reads
\begin{equation}
    \xi_{\text{PTA}}(T) \approx
    \left\{
       \begin{array}{lll}
          d_1 \, T^{-\frac{n-2}{n+2}},  & \,\,\,\,\,\,\,\, (T \ll 1);  \\ 
          \\
          1/ (a_n \ln T + b_n ),   & \,\,\,\,\,\,\,\, (T  \gg 1).
     \end{array} \right.       \label{xi_pta}
\end{equation}
Here, $d_1 = 4K \left[16(n-1)!! K \gamma \right]^{-2/(n+2)}$, $a_n = 1-2/n$, and $b_n = (2/n) \ln{ \left[ (n-1)!! \gamma \right] }- \ln{K}$. Compared to CVM result Eqs.(\ref{pow_low})-(\ref{pow_h2}), the PTA expression has correct form and correct $a_n$. But the coefficient $d_1$ and $b_n$ are inaccurate. For other quantities such as $\langle (z_1-z_2)^2 \rangle$ and correlation length $\xi(T)$, PTA also gives qualitatively correct asymptotic behavior.

\begin{figure}
 \vspace*{-3.0cm}
\begin{center}
  \includegraphics[width=380pt, height=340pt, angle=0]{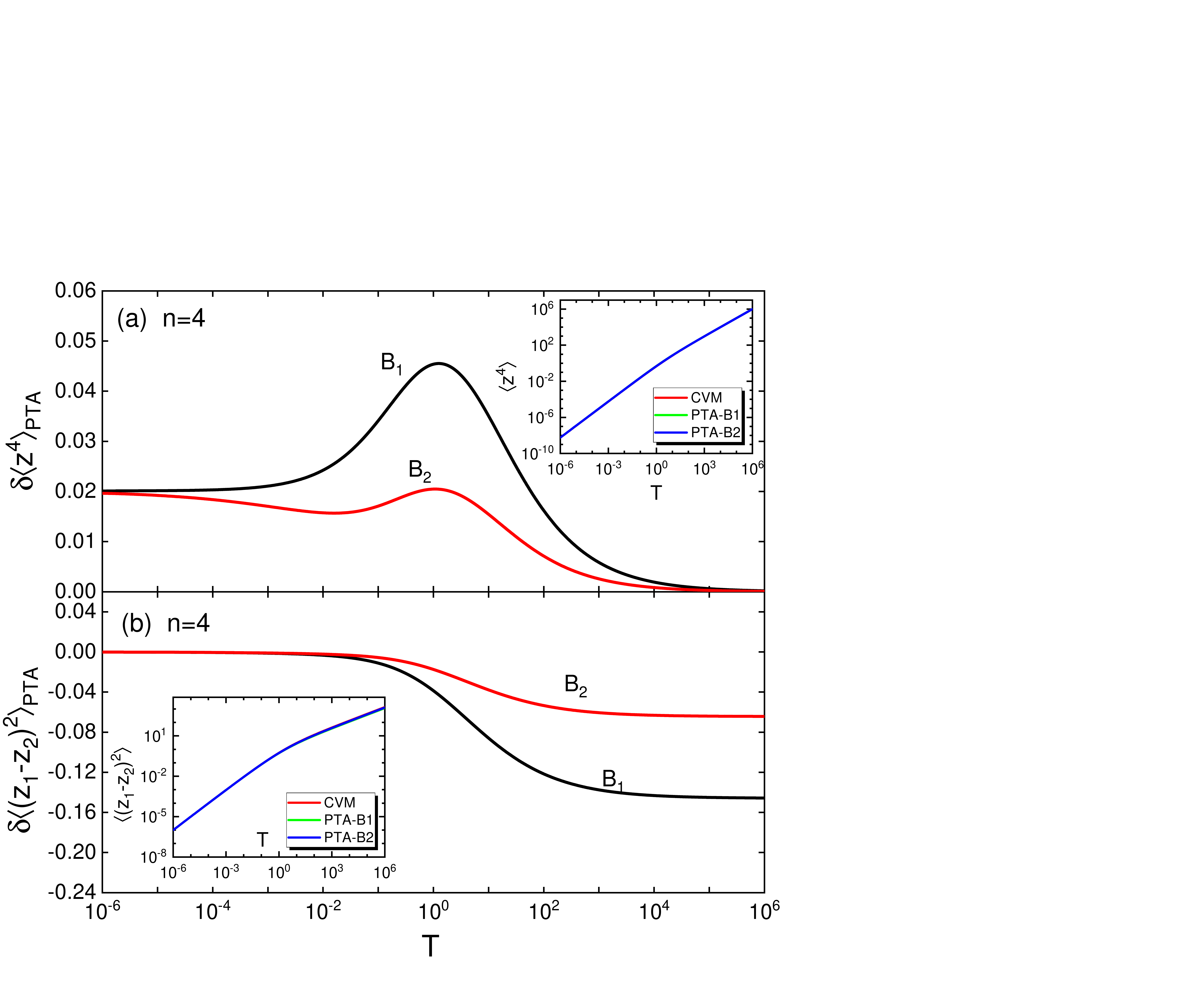}
  \vspace*{-1.5cm}
\end{center} \caption{(color online) Relative errors of PTA-$B_1$ and $B_2$ results with respect to CVM result for $n=4$. In (a), $\delta \langle z^4\rangle_{\text{PTA}} = \left( \langle z^4\rangle_{\text{PTA}}-\langle z^4\rangle_{\text{CVM}} \right)/ \langle z^4\rangle_{\text{CVM}}$ is plotted as functions of $T$. In (b), $\delta \langle (z_1-z_2)^2 \rangle_{\text{PTA}} = \left[ \langle (z_1-z_2)^2\rangle_{\text{PTA}}-\langle (z_1-z_2)^2\rangle_{\text{CVM}} \right] / \langle (z_1-z_2)^2\rangle_{\text{CVM}}$ is plotted as functions of $T$. In both (a) and (b), the black and read curves are from $B_1$ and $B_2$ basis, respectively. In the insets of (a) and (b), $\langle z^4 \rangle$ and $\langle (z_1-z_2)^2 \rangle$ from different methods are shown as functions of $T$, respectively.
} \label{Fig13}
\end{figure}

\begin{figure}
 \vspace*{-3.0cm}
\begin{center}
  \includegraphics[width=350pt, height=320pt, angle=0]{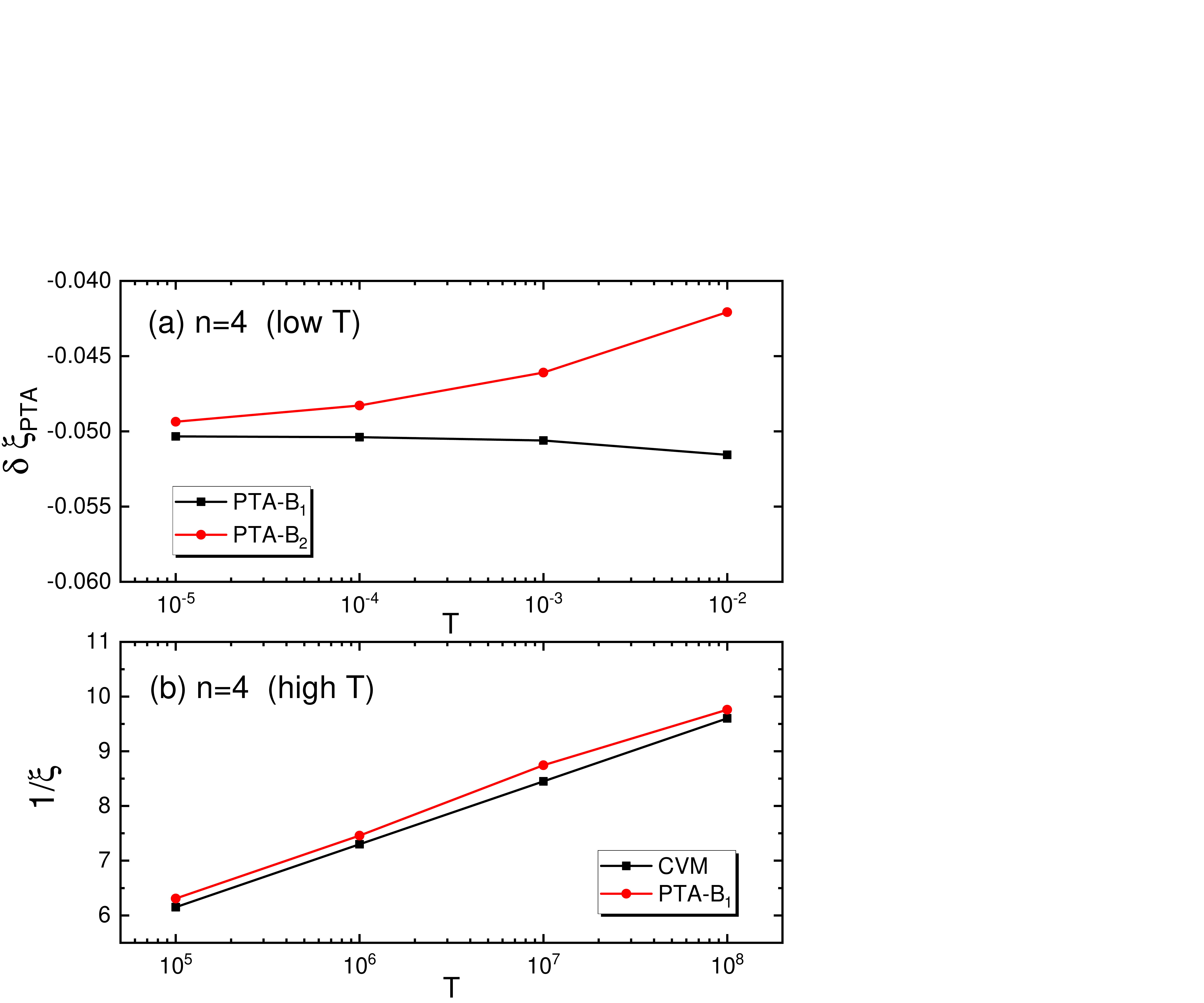}
  \vspace*{-1.0cm}
\end{center} \caption{(color online) (a) Relative errors $\delta \xi_{\text{PTA}}$ of the correlation length, obtained from PTA-$B_1$ and $B_2$, with respect to CVM result at low temperature. Here, $\delta \xi_{\text{PTA}} = (\xi_{\text{PTA}} - \xi_{\text{CVM}})/\xi_{\text{CVM}}$. (b) Comparison of $1/\xi$ from PTA-$B_1$ and that from CVM at high temperature. Both (a) and (b) are for $n=4$.
} \label{Fig14}
\end{figure}

Given the qualitative correctness of PTA results, below we make quantitative comparison between PTA and CVM. To be specific, we use the case of $n=4$ with $K=1$ and $\gamma=1$, {\it i.e.}, the $\phi^4$ model that has been studied by PTA on two successively larger basis, $B_1 = \{Q_k \}$ and $B_2 = \{Q_k, R_k\}$ \cite{PTAphi4}. We compare quantities $\langle z^4\rangle$, $\langle (z_1 - z_2)^2\rangle$ (Fig.\ref{Fig13}), and the correlation length $\xi(T)$ (Fig.\ref{Fig14}).

In Fig.\ref{Fig13}(a) and (b), we respectively plot the relative errors in the averages $\langle z^4\rangle_{\text{PTA}}$ and $\langle (z_1 - z_2)^2\rangle_{\text{PTA}}$ with respect to CVM results as functions of temperature. It is seen that in the temperature regime from $10^{-6}$ to $10^6$, PTA-$B_2$ result always has smaller error than that of PTA-$B_1$. The relative error in $\langle z^4\rangle_{\text{PTA}}$ on $B_1$ basis is smaller than $5\%$ in all temperature and tends to zero in the large $T$ limit. The relative errors in $\langle (z_1 - z_2)^2\rangle_{\text{PTA}}$ on $B_1$ and $B_2$ bases are negative. Their absolute values are smaller than $15\%$ in all temperature and tend to zero in the limit $T=0$. The insets of Fig.\ref{Fig13}(a) and (b) show that the curves $\langle z^4\rangle$ and $\langle (z_1 - z_2)^2\rangle$ from CVM, PTA-$B_1$, and PTA-$B_2$ are indistinguishable to eyes on the scale of the plot.

Lastly, we compare the correlation length of $\phi^4$ model obtained by PTA-$B_1$, PTA-$B_2$, and CVM. Fig.\ref{Fig14}(a) shows the relative error of correlation length $\delta \xi_{\text{PTA}} = (\xi_{\text{PTA}} - \xi_{\text{CVM}})/\xi_{\text{CVM}}$ for PTA-$B_1$ and PTA-$B_2$ in the low temperature regime. Both are negative and their absolute values are on the level of $5\%$ for $T \leq 10^{-2}$. Again the relative error of PTA-$B_2$ has smaller absolute value than that of PTA-$B_1$ in all temperatures. This and similar observation in Fig.\ref{Fig13} show that the precision of PTA improves with increasing basis dimension. In Fig.\ref{Fig14}(b), the high temperature part of $1/\xi(T)$ obtained from PTA-$B_1$ is compared with that from CVM. Both are linear curves on the linear-log plot, being consistent with Eq.(\ref{pow_h}). This confirms the qualitative correctness of PTA-$B_1$ in the high temperature limit. The largest relative error of PTA-$B_1$ in this quantity is on the order of $20\%$.  

Overall, making use of the exact CVM results, we gauge the accuracy of PTA-$B_1$ and -$B_2$ results. The comparison shows that both PTA-$B_1$ and -$B_2$ give correct low and high temperature asymptotic power laws. The quantitative accuracy is improved systematically by going from $B_1$ basis to $B_2$.

\section{Summary and Discussion}

In this paper, we implemented the CVM and LRT calculation for one dimensional KG lattice model, which has continuous degrees of freedom. We presented results for the single-site reduced density matrix $\rho^{(1)}(z)$, on-site averages $\langle |z^n|\rangle$ and $\langle z^2 \rangle$, nearest-neighbor coupling $\langle (z_1-z_2)^2\rangle$, specific heat $C_v$, and correlation  length $\xi$. We obtained their low and high temperature asymptotic scaling behaviors from both numerical and analytical analaysis. Exact expression for the asymptotic power law exponents are obtained. Using $\phi^4$ model (KG model at $n=4$), we compared the PTA results with CVM and established that PTA on $B_1$ and $B_2$ bases gave qualitatively correct and quantitatively well-controlled results. 

The CVM used in this work, i.e., CVM on the cluster of nearest-neighbor pairs and LRT, is exact for any one-dimensional lattice model with nearest-neighbor coupling and open boundary condition. Therefore, similar study can be carried out for a large class of models of this type, including Frenkel-Kontorova model \cite{Gillan1}, ding-a-ling model \cite{Casati1}, and models with more complicated local potentials such as the double-well potential \cite{Onodera1}, {\it etc.}. For systems with longer range interactions, exact thermodynamical properties can also be obtained from CVM on larger clusters if only the size of the cluster can cover the interaction range.

However, it is noted the present CVM has difficulty in studying the one-dimensional lattice models that conserve the total momentum, such as the FPU-$\beta$ model \cite{FPU1955}, Toda model \cite{Toda1,Toda2}, and Lennard-Jones chain model \cite{Mareschal1}, {\it etc.}. This is because for these models, the continuous $x_i$-translation symmetry of the system makes the $x_i$'s unconfined and hence the single-site reduced density $\rho^{(1)}(z)$ exactly zero. The formalism of CVM, which is based on $\rho^{(1)}(z)$, must be modified before it can be applied to such lattice models. This issue will be studied in the future.

It is natural to ask whether the exactness of CVM can be extended to time domain of the classical dynamical systems to study the interested relaxation or response properties. An extension of CVM to time domain is the path probability method \cite{Lopez}. It studies the kinetic process of the Ising-like lattice models and can be used to describe non-equilibrium time evolution process in alloy systems. However, a similar theory for describing the Hamilton dynamics of atoms has not been developed yet. Further study in this direction will be interesting.

\begin{acknowledgments}
   This work is supported by NSFC (Grant No.11974420).
\end{acknowledgments}

\appendix{}

\section{High Temperature Analysis of CVM and LRT Solution for $n>2$}

The CVM equations Eqs.(\ref{cvm_eq}) can be simplified into an integral equation for $Q(z)$ as
\begin{equation}   \label{A1}
   e^{\beta Q(z)} = \int_{-\infty}^{+\infty} dw \, e^{\beta \left[-V(z-w) - U(w) +Q(w) + \lambda_1 + \lambda_4 \right]} .
\end{equation}
For $n>2$, the local term $U(w) \sim |w|^n$ exceeds the non-local term $V(z-w) \sim (z-w)^2$ in the large $w$ regime. As a result, at high temperatures, the inter-site coupling $V(z-w)$ plays a minor role and we can expand the equations into perturbation series of $\beta V(z-w)$. In zero-th order of $\beta V(z-w)$, one gets $Q_{0}(z) = Q_0$, being a constant. To get the first-order correction, one put $Q_0(z)=Q_0$ into the right-hand side of the above equation. Expanding the right-hand side of Eq.(A1) to first-order of $\beta V(z-w)$ and we obtain
\begin{eqnarray}    \label{A2}
   && e^{\beta Q(z)}   \nonumber \\
   && \approx  e^{\beta(\lambda_1 + \lambda_4 + Q_0)} \int_{-\infty}^{+\infty} dw \, e^{-\beta U(w)}\left[ 1 - \beta V(z-w)  \right]      \nonumber \\
   && = e^{\beta(\lambda_1 + \lambda_4 + Q_0)} \left[ \int_{-\infty}^{+\infty} dw  \, e^{-\beta U(w)} \right] \left[1- \beta \langle V(z-w) \rangle_{\text{at}} \right]   \nonumber \\
   && \approx e^{\beta(\lambda_1 + \lambda_4 + Q_0)} \left[ \int_{-\infty}^{+\infty} dw \,  e^{-\beta U(w)} \right] e^{- \beta \langle V(z-w) \rangle_{\text{at}} }.
\end{eqnarray}
Here, 
\begin{eqnarray}
  \langle V(z-w) \rangle_{\text{at}} \equiv \frac{\int_{-\infty}^{+\infty} dw  \,  e^{-\beta U(w)} V(z-w) }{\int_{-\infty}^{+\infty} dw  \, e^{-\beta U(w)}}
\end{eqnarray}
is the average of $V(z-w)$ in the atomic limit. Inserting Eq.(\ref{A2}) into Eqs.(\ref{rho1_a}) and (\ref{cvm_eq}), one then gets the high temperature asymptotic form for $\rho^{(1)}(z)$ as
\begin{equation}
  \rho^{(1)}(z, T\rightarrow \infty) = c e^{-\beta \left[ U(z) + 2 \langle V(z-w) \rangle_{\text{at}} \right]} .
\end{equation}
Using $V(z-w) = (1/2) (z-w)^2$, one obtains
\begin{equation}
  \rho^{(1)}(z, T\rightarrow \infty) = \rho^{(1)}(0) e^{-\beta \left( \frac{1}{n}|z|^{n} + z^2  \right) } .
\end{equation}
Using the sum rule $\int_{-\infty}^{+\infty} \rho^{(1)}(z) dz = 1$, we obtain
\begin{eqnarray}
 1 &=& \rho^{(1)}(0)  \int_{-\infty}^{+\infty} dz \, e^{-\beta \left( \frac{1}{n}|z|^{n} + z^2  \right) }   \nonumber \\
  &\approx & \rho^{(1)}(0)  \int_{-\infty}^{+\infty} dz \, e^{-\beta \left( \frac{1}{n}|z|^{n} \right) }     \nonumber \\
  &=& \rho^{(1)}(0) 2 n^{\frac{1-n}{n}} T^{\frac{1}{n}} \Gamma\left( {\frac{1}{n}} \right).
\end{eqnarray}
This gives Eq.(\ref{rhohighT}) in the main text,
\begin{equation}
  \rho^{(1)}(z, T\rightarrow \infty) = \rho_0 T^{-\frac{1}{n}} e^{- \frac{1}{T} \left( \frac{1}{n}|z|^{n} + z^2  \right) },
\end{equation}
with $\rho_0 \approx 1/\left[2 n^{(1-n)/n} \Gamma(1/n) \right]$. 

To derive the high temperature asymptotic expression for the correlation function, we also expand the LRT equations Eqs.(\ref{lrt_eq}) and (\ref{fwz}) to leading order of $\beta V(z-w)$. We obtain
\begin{equation}
    F_{2}(w,z) \approx e^{-\beta Q(z) }e^{\beta \lambda_2(w) } \left[ 1 - \beta V(w-z) \right].
\end{equation}
Inserting it into the expression for $\chi_{2k}(z)$ in Eq.(\ref{lrt_eq}) and employing the symmetry properties $\lambda_2(-w) = \lambda_2(w)$, $Q(-z) = Q(z)$, and $\chi_{2k}(-z) = -\chi_{2k}(z)$, we solve the equations for $\chi_{2k}(z)$ as
\begin{eqnarray}
 \chi_{2k}(z) \approx B(z) + \beta e^{-ik} m_{k} z e^{-\beta Q(z)}.
\end{eqnarray}
Here, $m_k$ is given by
\begin{equation}
  m_k  \equiv \frac{ \int_{-\infty}^{+\infty} dz \, B(z) z e^{\beta \lambda_2(z)}}{ 1 - \beta e^{-ik} \int_{-\infty}^{+\infty} dz \, z^2 e^{\beta [ \lambda_2(z) - Q(z) ] } }.
 \end{equation}
Inserting Eq.(A9) and (A10) into Eq.(\ref{ck}) and evaluating the integral in the high temperature limit, i.e. the atomic limit $\beta V(z-w) \approx 0$ with $Q(z) \approx Q_0$, we obtain the asymptotic expression for $C(k)$,
\begin{eqnarray}    \label{A11}
 C(k)(T \rightarrow \infty) &\approx& \langle z^2 \rangle_{\text{at}} + 2\beta  \langle z^2 \rangle_{\text{at}}^2 f_k  \nonumber \\
 & \approx & \frac{\langle z^2 \rangle_{\text{at}}}{1 - 2 \beta f_k \langle z^2 \rangle_{\text{at}}},
\end{eqnarray}
with $f_k$ given by
\begin{equation}
  f_k = \text{Re} \left[ \frac{e^{ik}}{1 - \beta \langle z^2 \rangle_{\text{at}} e^{ik} }   \right].
\end{equation}
In the derivation, $\chi_{3k}(z) = \chi_{2k}^{\ast}(z)$ has been used. 

In the above equations, $\langle z^2 \rangle_{\text{at}} \equiv \int dz \, z^2 e^{-\beta U(z)} / \int dz e^{-\beta U(z)}$ can be calculated as
\begin{equation}
\langle z^2 \rangle_{\text{at}} = \left(nT \right)^{2/n} \frac{\Gamma(3/n)}{\Gamma(1/n)}. 
\end{equation}
For $n > 2$, we take the limit $T \rightarrow \infty$ and reduce $f_k \approx \cos{k}$. So we obtain the high temperature expression of $C(k)$ for $n>2$ as
\begin{equation}
   C(k) (T \rightarrow \infty) \approx \frac{\langle z^2 \rangle_{\text{at}}}{1 - 2 \beta \langle z^2 \rangle_{\text{at}} \cos{k} }.
\end{equation}

Now in order to extract the correlation length from the above expression, let us suppose that $C_{1j}$ on the periodic chain has the following form,
\begin{equation}
 C^{\text{fit}}_{1j} = C_0 \left( e^{- \frac{j-1}{\xi}} + e^{- \frac{L+1-j}{\xi}} \right),  \,\,\,\,\,\,\,(1 \leq j \leq L).
\end{equation}
Its Fourier transformation gives
\begin{eqnarray}    \label{Ck_fit}
   C^{\text{fit}}(k) &=&  \sum_{j=1}^{L} C^{\text{fit}}_{1j} e^{ik (j-1)}    \nonumber \\
   &=&  C_0 \frac{\sinh(1/\xi) }{\cosh(1/\xi) - \cos{k} }.
\end{eqnarray}
Now we compare Eq.(A14) with (A16). Taking the limit $T \rightarrow \infty$, which implies $\xi \rightarrow 0$, we obtain the high temperature asymptotic expression $\xi(T \rightarrow \infty)$ for $n>2$, i.e., Eqs.(\ref{pow_h}) and (\ref{pow_h2}) in the main text.

\section{Analysis of PTA-$B_1$ Solution}

In this appendix, we analyze the PTA-$B_1$ results in the low and high temperature limits.
Our starting point is the self-consistent equation Eq.(\ref{x2-PTA}). First, in the low temperature limit, $\langle x^2 \rangle$ tends to zero and the summation in Eq.(\ref{x2-PTA}) will be dominated by the $k \rightarrow 0$ contribution where $\omega_0^{2}(k) = 2K \left [1-\cos(k) \right] \approx K k^2$. In the large $L$ limit, we therefore obtain the approximate expression
\begin{eqnarray}
   \langle x^2 \rangle_{\text{PTA}}( T \rightarrow 0) &\approx & \frac{T}{2 \pi K} \int_{0}^{2\pi} dk \frac{1}{k^2 + \lambda^2}   \nonumber \\
       &\approx & \frac{T}{4K \lambda},  
\end{eqnarray}
with $\lambda^2 = (n-1)!! (\gamma/K) \langle x^2 \rangle^{n/2-1}$. This equation gives the low temperature expression for $\langle x^2 \rangle$ as
\begin{equation}
   \langle x^2 \rangle_{\text{PTA}}(T \rightarrow 0) = a T^{\frac{4}{n+2}}.
\end{equation}
Here $a = \left[16K\gamma (n-1)!! \right]^{-2/(n+2)}$. Using $\langle x^{m} \rangle \approx (m-1)!! \langle x^2 \rangle^{m/2}$, we obtain Eq.(\ref{avezm_pta}) in the main text. The correlation function in Eq.(\ref{corr-PTA}) is simplified similarly in the low $T$ limit as
\begin{equation}
  C(k)_{\text{PTA}}(T \rightarrow 0) \approx \frac{1}{\beta K} \frac{1}{k^2 + \lambda^2}.
\end{equation}
Comparing this expression with the fitting formula Eq.(A16) in Appensix A in the $k \rightarrow 0$ limit, one obtains
\begin{equation}
   \xi_{\text{PTA}}(T \rightarrow 0 ) \approx  \frac{1}{\lambda } = b T^{- \frac{n-2}{n+2}},
\end{equation}
with $b = 4K \left[16(n-1)!! K \gamma \right]^{-2/(n+2)}$.

In the high temperature limit, $\langle x^2 \rangle$ tends to infinity and the summation in Eq.(\ref{x2-PTA}) will be dominated by the $\lambda^2$ term. We then have
\begin{eqnarray}
   \langle x^2 \rangle_{\text{PTA}}(T \rightarrow \infty) \approx \frac{T}{(n-1)!! \gamma \langle x^2 \rangle^{n/2-1}}
\end{eqnarray}
which, together with $\langle x^m \rangle = (m-1)!! \langle x^2 \rangle^{m/2}$ ($m$ even), gives the asymptotic expression
\begin{equation}    \label{B6}
   \langle x^m \rangle_{\text{PTA}} (T \rightarrow \infty) = c T^{\frac{m}{n}},
\end{equation}
with $c = (m-1)!! \left[ (n-1)!! \gamma \right]^{-m/n}$.

Comparing the correlation function $\langle Q_k^{\ast} Q_k \rangle$ in Eq.(\ref{Ck_pta}) with the fitting expression Eq.(\ref{Ck_fit}) and using Eq.(\ref{B6}), one obtains
\begin{equation}
  \xi_{\text{PTA}}(T \rightarrow \infty) = \frac{1}{a^{\text{PTA}}_n \ln{T} + b^{\text{PTA}}_n},
\end{equation}
with $a_n^{\text{PTA}} = 1 - 2/n$ and $b_n^{\text{PTA}} = (2/n) \ln [(n-1)!!\gamma] - \ln K$. We find that $a_n^{\text{PTA}}$ is identical to the exact value from CVM, while $b_n^{\text{PTA}}$ is different from the CVM expression in Eq.(\ref{pow_h2}).


\end{document}